\documentclass[10pt]{article}
\usepackage[textwidth = 450 pt, textheight = 630 pt]{geometry}
\usepackage{amssymb}
\usepackage{amsmath}
\usepackage{appendix}
\usepackage{pdfsync}

\usepackage{color}
\definecolor{MyDarkBlue}{rgb}{0.15,0.15,0.45}
\usepackage[linktocpage=true]{hyperref}
\hypersetup{
colorlinks=true,
citecolor=MyDarkBlue,
linkcolor=MyDarkBlue,
urlcolor=MyDarkBlue,
pdfauthor={Paul Richmond},
pdftitle={Higher Derivative BLG: Lagrangian and Supersymmetry Transformations},
pdfsubject={hep-th}
}

\usepackage[numbers,sort&compress]{natbib}
\usepackage{hypernat}

\newcommand\sfrac[2]{{\textstyle\frac{#1}{#2}}}
\newcommand{\eref}[1]{Eq.\,(\ref{#1})}

\begin{document}

\thispagestyle{empty}

\renewcommand{\thefootnote}{\fnsymbol{footnote}}

\rightline{KCL-MTH-12-09}

\allowdisplaybreaks

\begin{flushright}
\end{flushright}
\begin{flushright}
\end{flushright}
\begin{center}

{\LARGE {\sc Higher Derivative BLG: Lagrangian and Supersymmetry Transformations\\}}

\bigskip

{\sc Paul Richmond\footnote{E-mail address:
                                 \href{mailto:paul.richmond@kcl.ac.uk}{\tt paul.richmond@kcl.ac.uk}}} \\
{Department of Mathematics\\
King's College London\\
The Strand\\
London WC2R 2LS, UK\\}

\end{center}

\bigskip

\centerline{\sc Abstract}

\vspace{0.4truecm}
\begin{center}
\begin{minipage}[c]{390pt}{
\noindent Working to lowest non-trivial order in fermions, we consider the four-derivative order corrected Lagrangian and supersymmetry transformations of the Euclidean Bagger-Lambert-Gustavsson theory. By demonstrating supersymmetric invariance of the Lagrangian we determine all numerical coefficients in the system. In addition, the supersymmetry algebra is shown to close on the scalar and gauge fields. We also comment on the extension to Lorentzian and other non-Euclidean $\mathcal{N}=8$ 3-algebra theories. 
}

\end{minipage}
\end{center}    

\newpage

\renewcommand{\thefootnote}{\arabic{footnote}}
\setcounter{footnote}{0}

\setcounter{page}{1}

\makeatletter
\@addtoreset{equation}{section}
\makeatother
\renewcommand{\theequation}{\thesection.\arabic{equation}}

\tableofcontents

\section{\sl Introduction}

The bosonic effective action for a single M2-brane \cite{Bergshoeff:1987cm} in static gauge and in a flat background with zero flux, is given by the abelian DBI action
\begin{align}
S_{M2} =& - T_{M2} \int d^3 x \ \sqrt{ - {\rm det} ( \eta_{\mu \nu} + \partial_\mu x^I \partial_\nu x^I ) } \, .
\end{align}
The terms in the integral can be expanded as a power series in $(\partial x)^2$ that is, a higher derivative expansion. 
After canonically renormalising the eight scalars so that $X^I = x^I \sqrt{T_{M2}}$, the leading order and next to leading order terms in the expansion are
\begin{align}
\nonumber
S_{M2} =\phantom{+} &\int d^3 x \ - \frac{1}{2} \partial_\mu X^I \partial^\mu X^I \\
+ &\int d^3 x \ \frac{1}{T_{M2}} \left( + \frac{1}{4} \partial_\mu X^I \partial^\mu X^J \partial_\nu X^I \partial^\nu X^J - \frac{1}{8} \partial_\mu X^I \partial^\mu X^I \partial_\nu X^J \partial^\nu X^J \right) \label{M2_action} \\
\nonumber
+ &\ldots \, , \phantom{\int} 
\end{align}
where we have ignored a constant and the ellipsis denotes terms $\mathcal{O}\big( ( 1/T_{M2})^2 \big)$ and higher.

The generalisation of the fully supersymmetric leading order M2-brane action to multiple M2-branes was first constructed by Bagger and Lambert \cite{Bagger:2006sk}\cite{Bagger:2007jr}\cite{Bagger:2007vi} and independently by Gustavsson \cite{Gustavsson:2007vu}. The Bagger-Lambert-Gustavsson (BLG) theory of M2-branes is an $\mathcal{N}=8$ supersymmetric field theory which is invariant under an $SO(8)$  R-symmetry. The original formulation of the theory required the use of an algebraic structure called a Euclidean 3-algebra and is now known to describe, in some cases, two M2-branes \cite{Lambert:2008et}\cite{Distler:2008mk}\cite{Lambert:2010ji}. Wider classes of Lorentzian and other non-Euclidean 3-algebra theories exist in the literature \cite{Gomis:2008uv}-\cite{deMedeiros:2009hf}, however their status as multiple M2-brane theories is unclear. Subsequent research has shown that the appropriate generalisation of the leading order term to arbitrary numbers of M2-branes is given by the ABJM theory \cite{Aharony:2008ug}.

There have been several papers which aim to determine the next to leading order i.e.\ the $1/T_{M2}$ higher derivative corrections to multiple M2-branes. It is known \cite{Nicolai:2003bp}\cite{deWit:2003ja}\cite{deWit:2004yr} that in three dimensions a non-abelian 2-form is dual to a scalar field. In \cite{Ezhuthachan:2008ch} this dualisation was applied to 3D super-Yang-Mills (the effective worldvolume theory of multiple D2-branes) and it was shown that it could be re-written as an $SO(8)$ invariant Lorentzian 3-algebra theory. Three-dimensional SYM arises simply by the appropriate dimensional reduction of 10D SYM and the higher derivative corrections to this have been uniquely determined (including quartic fermions in the Lagrangian) by superspace considerations in \cite{Cederwall:2001bt}\cite{Cederwall:2001td} and independently in \cite{Bergshoeff:2001dc} by calculating open-string scattering amplitudes. The first higher derivative corrections to the 10D SYM Lagrangian and supersymmetry transformations arise at order $\alpha'^2$ and the same is true in the reduction to three-dimensions. In \cite{Alishahiha:2008rs} the authors applied the analysis of \cite{Ezhuthachan:2008ch} to the $\alpha'^2$ corrections of the 3D SYM Lagrangian. The resulting $SO(8)$ invariant, Lorentzian 3-algebra formulation features only 3-brackets and covariant derivatives of the scalar and fermion fields. This lead to the conjecture that higher derivative corrections to the Euclidean BLG theory would be structurally identical to the Lorentzian theory and only feature 3-brackets and covariant derivatives.

A different approach was considered in \cite{Ezhuthachan:2009sr}. Here the most general $1/T_{M2}$ higher derivative M2-brane Lagrangian with arbitrary coefficients was considered. Then, using the `novel Higgs mechanism' \cite{Mukhi:2008ux} this was reduced uniquely to the four-derivative order correction of the D2-brane effective worldvolume theory. The results of \cite{Ezhuthachan:2009sr} applied both to the Euclidean BLG theory and Lorentzian 3-algebra theories and confirmed the conjecture of \cite{Alishahiha:2008rs}. Other attempts to construct the full non-linear action for multiple M2-branes include \cite{Li:2008ya}\cite{Pang:2008hw}\cite{Iengo:2008cq}.

The higher derivative corrected 3-algebra Lagrangians of \cite{Alishahiha:2008rs}\cite{Ezhuthachan:2009sr} are expected to possess maximal supersymmetry although this was not verified in either case. An attempt to determine next order corrections to the supersymmetry transformations was made in \cite{Low:2010ie}. Here Low applied the analysis of \cite{Ezhuthachan:2008ch} and \cite{Alishahiha:2008rs} at the level of the multiple D2-brane supersymmetry transformations. It was found that the $\alpha'^2$ corrections to the fermion supersymmetry could be written in an $SO(8)$ fashion but that the scalar transformation could not be. The gauge field supervariation was not considered. By taking an abelian truncation of the higher derivative Lorentzian 3-algebra action and showing it was supersymmetric, Low was able to partially determine the higher derivative scalar supersymmetry transformation. 

As it is not possible to derive higher derivative $SO(8)$ invariant 3-algebra valued supersymmetries from the multiple D2-brane ones by the 2-form/scalar dualisation approach, it seems the only way to unequivocally determine them is to examine the full supervariation of the higher derivative Lagrangian and by closing the superalgebra.  This is the approach we will take here, focussing solely on the Euclidean BLG theory of \cite{Ezhuthachan:2009sr}. 

The rest of the paper is as follows. In Section \ref{BLG_review} we will give a brief overview of the BLG model of M2-branes.
Our results are contained in Section \ref{HDer_BLG}; here we revisit the higher derivative action of \cite{Ezhuthachan:2009sr} and introduce our ansatz for the $1/T_{M2}$ corrections to the Euclidean BLG supersymmetry transformations. We determine all arbitrary coefficients in the system by examining the supervariation of the higher derivative Lagrangian for Euclidean BLG. In addition the supersymmetry algebra is shown to close on the scalar and gauge fields for the coefficients we find. 
Our conclusions can be found in Section \ref{conclusions} where we also comment on the extension to non-Euclidean theories and offer suggestions for further work. 
%

\section{\sl Overview of BLG} \label{BLG_review}
We begin with a brief overview\footnote{Detailed reviews of multiple M2-brane theories can be found in \cite{Bagger:2012jb} and \cite{Lambert:2012wr}.} of the BLG worldvolume theory of multiple M2-branes as presented in Ref.\ \cite{Bagger:2007jr}. As we have already noted, the BLG theory is a three-dimensional SCFT with $\mathcal{N}=8$ supersymmetry and an $SO(8)$ R\mbox{-}symmetry. Its field content consists of eight scalar fields, $X^I$, which parameterise directions transverse to the M2-brane worldvolume, fermions $\psi$ and a non-dynamical gauge field $A_\mu$. The fields take values in an $n$-dimensional real 3-algebra which we take to be spanned by a basis $T^a$,  $a=1, \ldots , n$. The 3-algebra is equipped with a totally anti-symmetric 3-bracket which defines the structure constants $f^{abc}{}_d$:
\begin{align}
[ T^a , T^b , T^c ] = f^{abc}{}_d T^d \, .
\end{align}
The structure constants inherit the total anti-symmetry of the 3-bracket so that
\begin{equation}
f^{abc}{}_d = f^{[abc]}{}_d \, . \label{f_anti_symmetry}
\end{equation}
The 3-bracket generates a gauge symmetry whose action on an arbitrary 3-algebra element $Y= Y_a T^a$ is 
\begin{equation}
\delta Y = [ \alpha , \beta , Y ] \, , \label{gauge_trans}
\end{equation}
where $\alpha$ and $\beta$ are two other elements of the 3-algebra. Requiring that this gauge symmetry acts as a derivation leads to the fundamental identity
\begin{equation}
f^{efg}{}_d f^{abc}{}_g = f^{efa}{}_g f^{bcg}{}_d + f^{efb}{}_g f^{cag}{}_d + f^{efc}{}_g f^{abg}{}_d \, \quad \text{i.e.} \quad f^{ [ abc}{}_g f^{ d ] efg}=0 \label{FI} \, .
\end{equation}
There is also an inner product on the 3-algebra which is symmetric and linear in both its entries and acts as a metric on the gauge indices. It is defined by
\begin{align}
h^{ab} = {\rm Tr} ( T^a T^b ) \, .
\end{align}
For any two 3-algebra elements $Y$ and $Z$ their inner product ${\rm Tr} (YZ)$ is required to be invariant under the gauge transformation \eqref{gauge_trans}. In basis form this leads to
\begin{align}
{\rm Tr} ( [ T^a , T^b , T^c ] T^d ) = - {\rm Tr} ( T^a [ T^b , T^c , T^d ] ) \ \ \text{ i.e.} \ \  h^{e(d} f^{a)bc}{}_e =  0 \, ,
\end{align}
which together with \eref{f_anti_symmetry} implies $f^{[abcd]} =0$. 

Real 3-algebras may to classified according to the signature of $h^{ab}$. When the inner product has one or more time-like directions there exist infinite families of 3-algebras. These are the Lorentzian 3-algebras of \cite{Gomis:2008uv}\cite{Benvenuti:2008bt}\cite{Ho:2008ei} and the multiple time-like 3-algebras of \cite{deMedeiros:2008bf}\cite{Ho:2009nk}\cite{deMedeiros:2009hf}.
For Euclidean signature where $h^{ab} \propto \delta^{ab}$, there is a unique (up to direct sums) finite-dimensional 3-algebra \cite{nagy-2007}\cite{Papadopoulos:2008sk}\cite{Gauntlett:2008uf}. In this                  case the structure constants are given by \cite{Bagger:2007jr}
\begin{equation}
f^{abcd} = \sfrac{2\pi}{k} \varepsilon^{abcd} \, ,
\end{equation}
with $a,b,c,d \in \{ 1 , 2 , 3 , 4 \}$ and $k$ is the integer Chern-Simons level \cite{Bagger:2007vi}. Hence the gauge algebra is $su(2) \times su(2) = so(4)$. Whilst there is a single 3-algebra and Lagrangian associated with the Euclidean theory there are two inequivalent gauge groups given by either $SU(2) \times SU(2) = Spin(4)$ or $(SU(2) \times SU(2)) / \mathbb{Z}_2 = SO(4)$ \cite{Lambert:2010ji}. We will refer to this 3-algebra as $\mathcal{A}_4$ and the theory as the $\mathcal{A}_4$ or Euclidean BLG theory. 
Investigation of the moduli space of the Euclidean BLG theory \cite{Lambert:2008et}\cite{Distler:2008mk}\cite{Lambert:2010ji} identifies it as the worldvolume theory for a pair of M2-branes propagating in an orbifold characterised by the level $k$. 

The lowest order multiple M2-brane supersymmetry transformations are,
\begin{align}
\delta X^I_a &= i \bar{\epsilon} \Gamma^I \psi_a \, , \label{BLG_susy_s} \\[6pt]
\delta \tilde{A}_{\mu}{}^b{}_a &= i \bar{\epsilon} \Gamma_\mu \Gamma_I X^I_c \psi_d f^{cdb}{}_a \, , \label{BLG_susy_g} \\[6pt]
\delta \psi_a &= \Gamma^\mu \Gamma^I \epsilon D_\mu X^I_a - \sfrac{1}{6} \Gamma^{IJK} \epsilon X^I_b X^J_c X^K_d f^{cdb}{}_a \, , \label{BLG_susy_f}
\end{align}
where $\tilde{A}_{\mu}{}^b{}_a = A_{\mu \, cd} f^{cdb}{}_a$ and $D_\mu X^I_a = \partial_\mu X^I_a - \tilde{A}_{\mu}{}^b{}_a X^I_b$. The commutator of two supersymmetries on the fields gives
\begin{align}
[\delta_1,\delta_2] X^I_a =& - 2 i ( \bar{\epsilon}_2 \Gamma^\mu \epsilon_1 ) D_\mu X^I_a - i ( \bar{\epsilon}_2 \Gamma_{JK} \epsilon_1 ) X^J_c X^K_d X^I_b f^{cdb}{}_a \, , \\[8pt]
%
%
\nonumber
[\delta_1,\delta_2] \tilde A_\mu{}^b{}_a =& + 2 i ( \bar{\epsilon}_2 \Gamma^\nu \epsilon_1 ) \varepsilon_{\mu\nu\lambda} \left(X^I_c D^\lambda X^I_d +\sfrac{i}{2}\bar\psi_c\Gamma^\lambda\psi_d \right) f^{cdb}{}_{a} \\
&- 2i ( \bar{\epsilon}_2 \Gamma_{IJ} \epsilon_1 ) X^I_c D_\mu X^J_d f^{cdb}{}_a \, , \\[8pt]
%
%
\nonumber
[\delta_1,\delta_2] \psi_a  =& - 2 i ( \bar{\epsilon}_2 \Gamma^\mu \epsilon_1 ) D_\mu \psi_a - i ( \bar{\epsilon}_2 \Gamma_{IJ} \epsilon_1 ) X^I_c X^J_d \psi_b f^{cdb}{}_a \\
\nonumber
&+ i(\bar\epsilon_2\Gamma_\nu\epsilon_1)\Gamma^\nu\left(\Gamma^\mu D_\mu\psi_a +\sfrac{1}{2}\Gamma_{IJ} X^I_c X^J_d \psi_b f^{cdb}{}_a \right) \\
&- \sfrac{i}{4}(\bar\epsilon_2\Gamma_{KL}\epsilon_1)\Gamma^{KL}\left(\Gamma^\mu D_\mu\psi_a +\sfrac{1}{2}\Gamma_{IJ} X^I_c X^J_d \psi_b f^{cdb}{}_a \right) \, .
\end{align}
Hence the supersymmetries close on to translations and gauge transformations after imposing the following equations of motion
\begin{align}\label{EOMS}
E_{A_\lambda{}^a{}_b} &= \sfrac{1}{2} \varepsilon^{\mu\nu\lambda} \tilde F_{\mu\nu}{}^b{}_a - \left(X^I_c D^\lambda X^I_d +\sfrac{i}{2}\bar\psi_c\Gamma^\lambda\psi_d \right)f^{cdb}{}_{a} = 0 \, , \\[8pt]
E_{\bar{\psi}^a} &= \Gamma^\mu D_\mu \psi_a +\sfrac{1}{2}\Gamma_{IJ}X^I_cX^J_d \psi_bf^{cdb}{}_{a} =0 \, , 
\end{align}
where $\tilde F_{\mu\nu}{}^b{}_a = \partial_\nu \tilde{A}_\mu{}^b{}_a - \partial_\mu \tilde{A}_\nu{}^b{}_a + \tilde{A}_\nu{}^b{}_c \tilde{A}_\mu{}^c{}_a - \tilde{A}_\mu{}^b{}_c \tilde{A}_\nu{}^c{}_a$ is a gauge field strength. The scalar equation of motion:
\begin{equation}
E_{X^{Ia}} = D^2X^I_a-\sfrac{i}{2}\bar\psi_c\Gamma^{IJ} X^J_d\psi_b f^{cdb}{}_a +\sfrac{1}{2}f^{bcd}{}_{a}f^{efg}{}_{d} X^J_bX^K_cX^I_eX^J_fX^K_g = 0 \, ,
\end{equation}
can be identified by taking the supervariation of the fermion equation of motion.
The supersymmetric Lagrangian which gives rise to these field equations is
\begin{align}\label{action}
\nonumber 
{\cal L}_{BLG} =& {\rm Tr} \left( -\sfrac{1}{2} D_\mu X^I D^\mu X^I - \sfrac{1}{12} [X^I,X^J,X^K] [X^I,X^J,X^K] +\sfrac{i}{2}\bar\psi\Gamma^\mu D_\mu \psi + \sfrac{i}{4}\bar\psi \Gamma_{IJ} [ X^I , X^J , \psi ] \right) \\[6pt]
&+\sfrac{1}{2}\varepsilon^{\mu\nu\lambda} \left(f^{abcd}A_{\mu ab}\partial_\nu A_{\lambda cd} +\sfrac{2}{3}f^{cda}{}_gf^{efgb} A_{\mu ab}A_{\nu cd}A_{\lambda ef} \right) \, .
\end{align}
%
\section{\sl Higher Derivative Lagrangian and Supersymmetries} \label{HDer_BLG}

We begin with the most general four-derivative order Lagrangian as considered in \cite{Ezhuthachan:2009sr}, to lowest non-trivial order in fermions
\begin{align}
\nonumber
\mathcal{L}_{T^{-1}_{M2}} = \sfrac{1}{T_{M2}} {\rm STr} \Big\{ &+ \bold{a} \, D^\mu X^I D_\mu X^J  D^\nu X^J D_\nu X^I + \bold{b} \, D^\mu X^I D_\mu X^I D^\nu X^J D_\nu X^J \\
\nonumber
&+ \bold{c} \, \varepsilon^{\mu\nu\lambda}\, X^{IJK} D_\mu X^I D_\nu X^J D_\lambda X^K \\
\nonumber
&+ \bold{d} \, X^{IJK} X^{IJL} D^\mu X^K D_\mu X^L + \bold{e} \, X^{IJK} X^{IJK} D^\mu X^L  D_\mu X^L \\
\nonumber
&+ \bold{f} \, X^{IJK} X^{IJK} X^{LMN} X^{LMN} + \bold{g} \, X^{IJK} X^{IJL} X^{KMN} X^{LMN} \\
%
%
\nonumber
&+ i \hat{\bold{d}} \, \bar{\psi} \Gamma^\mu \Gamma^{IJ} D^\nu\psi {D}_\mu X^{I} {D}_\nu X^J + i \hat{\bold{e}} \, \bar{\psi}\Gamma^\mu D^\nu\psi  {D}_\mu X^{I} {D}_\nu X^I \\
\nonumber
&+ i \hat{\bold{f}} \, \bar{\psi}\Gamma^{IJKL} D^\nu\psi\; X^{IJK}  {D}_\nu X^L + i \hat{\bold{g}} \, \bar{\psi}\Gamma^{IJ}  D^\nu\psi\; X^{IJK}{D}_\nu X^K \\
\nonumber
&+ i \hat{\bold{h}} \, \bar{\psi}\Gamma^{IJ}[X^J,X^{K},\psi] {D}^\mu X^{I} {D}_\mu X^K \\
\nonumber
&+ i \hat{\bold{i}} \, \bar{\psi}\Gamma^{\mu\nu}[X^I,X^{J},\psi] {D}_\mu X^{I}{D}_\nu X^J + i \hat{\bold{j}} \, \bar{\psi}\Gamma^{\mu\nu}\Gamma^{IJ}[X^J,X^{K},\psi] {D}_\mu X^{I}{D}_\nu X^K \\
\nonumber
&+ i \hat{\bold{k}} \, \bar{\psi}\Gamma^\mu\Gamma^{IJ}[X^K,X^{L},\psi] {D}_\mu X^{I}X^{JKL} + i \hat{\bold{l}} \, \bar{\psi}\Gamma^\mu[X^I,X^{J},\psi] {D}_\mu X^{K}X^{IJK} \\
\nonumber
&+ i \hat{\bold{m}} \, \bar{\psi}\Gamma^\mu\Gamma^{IJKL}[X^L,X^M,\psi] {D}_\mu X^M X^{IJK} + i \hat{\bold{n}} \, \bar{\psi}\Gamma^\mu\Gamma^{IJ}[X^K,X^{L},\psi] {D}_\mu X^L X^{IJK} \\
&+ i \hat{\bold{o}} \, \bar{\psi}\Gamma^{IJKL}[X^M,X^N,\psi] X^{IJL}X^{KMN} + i \hat{\bold{p}} \, \bar{\psi}\Gamma^{IJ}[X^K,X^{L},\psi] X^{IJM}X^{KLM} \Big\} \, . \label{L_Higher_Starting_Point}
\end{align}
We have adopted the notation $X^{IJK} := [ X^I , X^J , X^K ]$ which we will use to save space wherever possible. Let us make some comments on this Lagrangian. The symmetrised trace of four basis elements of the $\mathcal{A}_4$ 3-algebra is given by ${\rm STr} \big\{ T^a T^b T^c T^d \big\} = d^{abcd}$ and is totally symmetric and linear in its four entries. Next, we require that each term within this higher derivative Lagrangian is gauge invariant. Acting on a generic four-derivative order term with the gauge transformation \eqref{gauge_trans} we see that this requirement leads to 
\begin{align}
{\rm STr} \big\{ [ \alpha , \beta, Y_1 ] Y_2 Y_3 Y_4 + Y_1 [ \alpha , \beta, Y_2 ] Y_3 Y_4 + Y_1 Y_2 [ \alpha , \beta, Y_3 ] Y_4 + Y_1 Y_2 Y_3 [ \alpha , \beta, Y_4 ] \big\} =0 \, , \label{Gauge_Inv}
\end{align}
where $Y_1, \ldots, Y_4$ are arbitrary fields. In basis form this symmetrised trace invariance condition reads
\begin{equation}
d^{abcd} f^{efg}{}_a + d^{aecd} f^{bfg}{}_a + d^{abed} f^{cfg}{}_a + d^{abce} f^{dfg}{}_a =0 \ \ \text{ i.e.} \ \ d^{a(bcd} f^{e)fg}{}_a = 0 \, , \label{Gauge_Inv_basis}
\end{equation}
and can be seen as a generalisation of the trace invariance property: $ h^{a(b} f^{e)fg}{}_a = 0$.

There are further identities we can construct using the symmetrised trace. To start with we note that due to their simple nature the structure constants of the $\mathcal{A}_4$ 3-algebra satisfy 
\begin{equation}
f^{[abcd} f^{e] fgh} =0 \ \ \text{ i.e.} \ \ f^{abcd} f^{efgh} = + f^{bced} f^{afgh} - f^{cead} f^{bfgh} + f^{eabd} f^{cfgh} - f^{eabc} f^{dfgh} \, . \label{A4_Id}
\end{equation}
We can combine this identity with the symmetrised trace to find
\begin{align}
{\rm STr} \big\{ T_d T_h T_i T_j \big\} f^{abcd} f^{efgh} =& {\rm STr} \big\{ T_d T_h T_i T_j \big\} ( f^{bced} f^{afgh} - f^{cead} f^{bfgh} + f^{eabd} f^{cfgh} ) \, ,
\end{align}
where the final term, ${\rm STr} \big\{ T_d T_h T_i T_j \big\} f^{eabc} f^{dfgh} $, vanishes because of symmetry/anti-symmetry under $d \leftrightarrow h$. Contracting the gauge indices with the fields leads to the following identities
\begin{align}
{\rm STr} \Big\{ \alpha \beta \, X^{I_1 I_2 I_3} X^{J_1 J_2 J_3} \Big\} = {\rm STr} \Big\{ \alpha \beta \left(  X^{J_1 I_2 I_3} X^{I_1 J_2 J_3} + X^{I_1 J_1 I_3} X^{I_2 J_2 J_3} + X^{I_1 I_2 J_1} X^{I_3 J_2 J_3} \right) \Big\} \, , \label{Useful_Id} 
\end{align}
\begin{align}
\nonumber
{\rm STr} \Big\{ \alpha \beta \, [ X^{I_1} , X^{I_2} , \gamma ] X^{J_1 J_2 J_3} \Big\} = {\rm STr} \Big\{ \alpha \beta \, \big( [ X^{J_1} , X^{I_2} , \gamma ] X^{I_1 J_2 J_3} + &[ X^{I_1} , X^{J_1} , \gamma ] X^{I_2 J_2 J_3} \\
+ &[ X^{J_2} , X^{J_3} , \gamma ] X^{I_1 I_2 J_1} \big) \Big\} \, , \label{Useful_Id2}
\end{align}
where $\alpha$, $\beta$ and $\gamma$ are arbitrary fields and $I_1 , J_1 , \ldots$ are transverse Lorentz indices. 

The starting ansatz for the four-derivative order Lagrangian can be simplified using these identities. Equation \eqref{Useful_Id} shows that the $\bold{f}$ and $\bold{g}$ terms in $\mathcal{L}_{1/T_{M2}}$ are proportional to each other. The same equation, together with anti-symmetry in the $\Gamma$-matrix indices, tells us that the term in $\mathcal{L}_{1/T_{M2}}$ with coefficient $\hat{\bold{o}}$ is identically zero. Similarly, the term with coefficient $\hat{\bold{m}}$ is identically zero through the use of \eref{Useful_Id2}. We subsequently drop the terms with coefficients $\bold{g}$, $\hat{\bold{m}}$ and $\hat{\bold{o}}$ to leave
\begin{align}
\nonumber
\mathcal{L}_{T^{-1}_{M2}} = \sfrac{1}{T_{M2}} {\rm STr} \Big\{ &+ \bold{a} \, D^\mu X^I D_\mu X^J  D^\nu X^J D_\nu X^I + \bold{b} \, D^\mu X^I D_\mu X^I D^\nu X^J D_\nu X^J \\
\nonumber
&+ \bold{c} \, \varepsilon^{\mu\nu\lambda}\, X^{IJK} D_\mu X^I D_\nu X^J D_\lambda X^K \\
\nonumber
&+ \bold{d} \, X^{IJK} X^{IJL} D^\mu X^K D_\mu X^L + \bold{e} \, X^{IJK} X^{IJK} D^\mu X^L  D_\mu X^L \\
\nonumber
&+ \bold{f} \, X^{IJK} X^{IJK} X^{LMN} X^{LMN} \\
%
%
\nonumber
&+ i \hat{\bold{d}} \, \bar{\psi} \Gamma^\mu \Gamma^{IJ} D^\nu\psi {D}_\mu X^{I} {D}_\nu X^J + i \hat{\bold{e}} \, \bar{\psi}\Gamma^\mu D^\nu\psi  {D}_\mu X^{I} {D}_\nu X^I \\
\nonumber
&+ i \hat{\bold{f}} \, \bar{\psi}\Gamma^{IJKL} D^\nu\psi\; X^{IJK}  {D}_\nu X^L + i \hat{\bold{g}} \, \bar{\psi}\Gamma^{IJ}  D^\nu\psi\; X^{IJK}{D}_\nu X^K \\
\nonumber
&+ i \hat{\bold{h}} \, \bar{\psi}\Gamma^{IJ}[X^J,X^{K},\psi] {D}^\mu X^{I} {D}_\mu X^K \\
\nonumber
&+ i \hat{\bold{i}} \, \bar{\psi}\Gamma^{\mu\nu}[X^I,X^{J},\psi] {D}_\mu X^{I}{D}_\nu X^J + i \hat{\bold{j}} \, \bar{\psi}\Gamma^{\mu\nu}\Gamma^{IJ}[X^J,X^{K},\psi] {D}_\mu X^{I}{D}_\nu X^K \\
\nonumber
&+ i \hat{\bold{k}} \, \bar{\psi}\Gamma^\mu\Gamma^{IJ}[X^K,X^{L},\psi] {D}_\mu X^{I}X^{JKL} + i \hat{\bold{l}} \, \bar{\psi}\Gamma^\mu[X^I,X^{J},\psi] {D}_\mu X^{K}X^{IJK} \\
\nonumber
&+ i \hat{\bold{n}} \, \bar{\psi}\Gamma^\mu\Gamma^{IJ}[X^K,X^{L},\psi] {D}_\mu X^L X^{IJK} \\
&+ i \hat{\bold{p}} \, \bar{\psi}\Gamma^{IJ}[X^K,X^{L},\psi] X^{IJM}X^{KLM} \Big\} . \label{L_Higher}
\end{align}

We now give the general starting point for the $1/T_{M2}$ higher derivative corrections to the $\mathcal{N}=8$ supersymmetry transformations which are consistent with mass dimension, 3-algebra index structure, parity under $\Gamma_{012}$ and Lorentz invariance. We assume that the higher derivative scalar and fermion supersymmetry transformations are built out of $\psi$, $DX$ and $[X,X,X]$ only. In particular, as the Chern-Simons term in $\mathcal{L}_{BLG}$ does not receive higher derivative corrections, the gauge field strength is not present in the $1/T_{M2}$ supersymmetries. The gauge field variation additionally requires the presence of a `bare' scalar field. 

Our ansatz for the scalar supersymmetry transformation, to lowest order in fermions, is
\begin{align}
\delta' X^I_a &= \sfrac{1}{T_{M2}} \left( \delta_{2  DX}' X^I_a + \delta_{1  DX}' X^I_a + \delta_{0  DX}' X^I_a \right) \, ,
\end{align}
where
\begin{align}
\nonumber
\delta_{2  DX}' X^I_a =&+ i s_1 ( \bar{\epsilon} \Gamma^{IJK} \Gamma^{\mu \nu} \psi_b ) D_{\mu} X^{J}_c D_{\nu} X^{K}_d \ d^{bcd}{}_{a} \\
\nonumber
&+  i s_2 ( \bar{\epsilon} \Gamma^{J} \Gamma^{\mu \nu} \psi_b ) D_{\mu} X^{I}_c D_{\nu} X^{J}_d \ d^{bcd}{}_{a} \\
\nonumber
&+  i s_3 ( \bar{\epsilon} \Gamma^{J} \psi_b ) D_{\mu} X^{I}_c D^{\mu} X^{J}_d \ d^{bcd}{}_{a} \\
&+  i s_4 ( \bar{\epsilon} \Gamma^{I} \psi_b ) D_{\mu} X^{J}_c D^{\mu} X^{J}_d \ d^{bcd}{}_{a} \label{S1} \, , \\ 
\nonumber \\
\nonumber
\delta_{1  DX}' X^I_a =&+ i s_5 ( \bar{\epsilon} \Gamma^{IJKLM} \Gamma^{\mu} \psi_b ) D_{\mu} X^{J}_c X^{KLM}_d \ d^{bcd}{}_{a} \\
\nonumber
&+ i s_6	( \bar{\epsilon} \Gamma^{KLM} \Gamma^{\mu} \psi_b ) D_{\mu} X^{I}_c X^{KLM}_d \ d^{bcd}{}_{a} \\
\nonumber
&+ i s_7	( \bar{\epsilon} \Gamma^{JLM} \Gamma^{\mu} \psi_b ) D_{\mu} X^{J}_c X^{ILM}_d \ d^{bcd}{}_{a} \\
\nonumber
&+ i s_8	( \bar{\epsilon} \Gamma^{ILM} \Gamma^{\mu} \psi_b ) D_{\mu} X^{J}_c X^{JLM}_d \ d^{bcd}{}_{a} \\
&+ i s_9	( \bar{\epsilon} \Gamma^{M} \Gamma^{\mu} \psi_b ) D_{\mu} X^{J}_c X^{IJM}_d \ d^{bcd}{}_{a} \label{S2} \, , \\
\nonumber \\
\nonumber
\delta_{0  DX}' X^I_a =&+ i s_{10} ( \bar{\epsilon} \Gamma^{I} \psi_b ) X^{JKL}_c X^{JKL}_d \ d^{bcd}{}_{a} \\
&+ i s_{11} ( \bar{\epsilon} \Gamma^{L} \psi_b ) X^{JKL}_c X^{JKI}_d \ d^{bcd}{}_{a} \label{S3} \, .
\end{align}
The ansatz for the fermion supersymmetry transformation is
\begin{align}
\delta' \psi_a &= \sfrac{1}{T_{M2}} \left( \delta_{3  DX}' \psi_a + \delta_{2  DX}' \psi_a + \delta_{1  DX}' \psi_a + \delta_{0  DX}' \psi_a \right) \, ,
\end{align}	
where
\begin{align}
\nonumber
\delta_{3  DX}' \psi_a =&+ f_1 \Gamma^{JKL} \Gamma^{\mu \nu \lambda} \epsilon D_{\mu} X^{J}_b D_{\nu} X^{K}_c D_{\lambda} X^{L}_d \ d^{bcd}{}_{a} \\
\nonumber
&+ f_2 \Gamma^{K} \Gamma^{\mu} \epsilon D_{\mu} X^{J}_b D_{\nu} X^{J}_c D^{\nu} X^{K}_d \ d^{bcd}{}_{a} \\
&+ f_3 \Gamma^{K} \Gamma^{\mu} \epsilon D_{\mu} X^{K}_b D_{\nu} X^{J}_c D^{\nu} X^{J}_d \ d^{bcd}{}_{a} \, , \\
\nonumber \\
\nonumber
\delta_{2 DX}' \psi_a =&+ f_4 \Gamma^{JKLMN} \Gamma^{\mu \nu} \epsilon  D_\mu X^J_b D_\nu X^K_c X^{LMN}_d \, d^{bcd}{}_a \\
\nonumber
&+ f_5 \Gamma^{KLM} \Gamma^{\mu \nu} \epsilon D_\mu X^J_b D_\nu X^K_c X^{JLM}_d \, d^{bcd}{}_a \\
\nonumber
&+ f_6 \Gamma^M \Gamma^{\mu \nu} \epsilon D_\mu X^J_b D_\nu X^K_c X^{JKM}_d \, d^{bcd}{}_a \\
\nonumber
&+ f_7 \Gamma^{KLM} \epsilon D_\mu X^J_b D^\mu X^J_c X^{KLM}_d \, d^{bcd}{}_a \\
&+ f_8 \Gamma^{KLM} \epsilon D_\mu X^J_b D^\mu X^K_c X^{JLM}_d \, d^{bcd}{}_a \, , \\
\nonumber \\
\nonumber
\delta_{1 DX}' \psi_a =&+ f_9 \Gamma^J \Gamma^\mu \epsilon D_\mu X^J_b X^{KLM}_c X^{KLM}_d \, d^{bcd}{}_a \\
&+ f_{10} \Gamma^M \Gamma^\mu \epsilon D_\mu X^J_b X^{JKL}_c X^{KLM}_d \, d^{bcd}{}_a \, , \\
\nonumber \\
\delta_{0 DX}' \psi_a =&+ f_{11} \Gamma^{NOP} \epsilon X^{JKL}_b X^{JKL}_c X^{NOP}_d \, d^{bcd}{}_a \, .
\end{align}
Finally, the ansatz for the gauge field variation, again to lowest order in fermions, is 
\begin{equation}
\delta' \tilde{A}_{\mu}{}^b{}_a = \sfrac{1}{T_{M2}} \left( \delta_{2  DX}' \tilde{A}_{\mu}{}^b{}_a + \delta_{1  DX}' \tilde{A}_{\mu}{}^b{}_a + \delta_{0  DX}' \tilde{A}_{\mu}{}^b{}_a \right) \, ,
\end{equation}
where
\begin{align}
\nonumber
\delta_{2  DX}' \tilde{A}_{\mu}{}^b{}_a =&+ i g_1 ( \bar{\epsilon} \Gamma_\mu \Gamma^I \psi_e ) D_\nu X^J_f D^\nu X^J_g X^I_c \, d^{efg}{}_d f^{cdb}{}_a \\
\nonumber
&+ i g_2 ( \bar{\epsilon} \Gamma^\nu \Gamma^I \psi_e ) D_\mu X^J_f D_\nu X^J_g X^I_c \, d^{efg}{}_d f^{cdb}{}_a \\
\nonumber
&+ i g_3 ( \bar{\epsilon} \Gamma^\nu \Gamma^J \psi_e ) D_\mu X^J_f D_\nu X^I_g X^I_c \, d^{efg}{}_d f^{cdb}{}_a \\
\nonumber
&+ i g_4 ( \bar{\epsilon} \Gamma^\nu \Gamma^J \psi_e ) D_\mu X^I_f D^\nu X^J_g X^I_c \, d^{efg}{}_d f^{cdb}{}_a \\
\nonumber
&+ i g_5 ( \bar{\epsilon} \Gamma_\mu \Gamma^J \psi_e ) D_\nu X^J_f D^\nu X^I_g X^I_c \, d^{efg}{}_d f^{cdb}{}_a \\
\nonumber
&+ i g_6 ( \bar{\epsilon} \Gamma_{\mu \nu \lambda} \Gamma^J \psi_e ) D^\nu X^J_f D^\lambda X^I_g X^I_c \, d^{efg}{}_d f^{cdb}{}_a \\
\nonumber
&+ i g_7 ( \bar{\epsilon} \Gamma_{\mu \nu \lambda} \Gamma^{IJK} \psi_e ) D^\nu X^J_f D^\lambda X^K_g X^I_c \, d^{efg}{}_d f^{cdb}{}_a \\
&+ i g_8 ( \bar{\epsilon} \Gamma^{\nu} \Gamma^{IJK} \psi_e ) D_\mu X^J_f D^\nu X^K_g X^I_c \, d^{efg}{}_d f^{cdb}{}_a \label{G1} \, , \\
\nonumber \\
\nonumber
\delta_{1  DX}' \tilde{A}_{\mu}{}^b{}_a =&+i g_9 ( \bar{\epsilon} \Gamma_{\mu \nu} \Gamma^{KLM} \psi_e ) D^\nu X^I_f X^{KLM}_g X^I_c \, d^{efg}{}_d f^{cdb}{}_a \\
\nonumber
&+i g_{10} ( \bar{\epsilon} \Gamma_{\mu \nu} \Gamma^{JLM} \psi_e ) D^\nu X^J_f X^{ILM}_g X^I_c \, d^{efg}{}_d f^{cdb}{}_a \\
\nonumber
&+i g_{11} ( \bar{\epsilon} \Gamma_{\mu \nu} \Gamma^{M} \psi_e ) D^\nu X^J_f X^{IJM}_g X^I_c \, d^{efg}{}_d f^{cdb}{}_a \\
\nonumber
&+i g_{12} ( \bar{\epsilon} \Gamma^{KLM} \psi_e ) D_\mu X^I_f X^{KLM}_g X^I_c \, d^{efg}{}_d f^{cdb}{}_a \\	
\nonumber
&+i g_{13} ( \bar{\epsilon} \Gamma^{JLM} \psi_e ) D_\mu X^J_f X^{ILM}_g X^I_c \, d^{efg}{}_d f^{cdb}{}_a \\	
&+i g_{14} ( \bar{\epsilon} \Gamma^{M} \psi_e ) D_\mu X^J_f X^{IJM}_g X^I_c \, d^{efg}{}_d f^{cdb}{}_a \label{G2} \, , \\
\nonumber \\
\delta_{0  DX}' \tilde{A}_{\mu}{}^b{}_a =&+ i g_{15} ( \bar{\epsilon} \Gamma_{\mu} \Gamma^{I} \psi_e ) X^{JKL}_f X^{JKL}_g X^I_c \, d^{efg}{}_d f^{cdb}{}_a \label{G3} \, .
\end{align}
There are other terms which are consistent with mass dimensions etc.\ that could be added to the $\delta'$ variations however, we can apply the $\mathcal{A}_4$ identity $f^{[abcd} f^{e]fgh} =0$ at the level of the supersymmetry transformations to find
\begin{equation}
\alpha_b X^{I_1 I_2 I_3}_c X^{J_1 J_2 J_3}_d \, d^{bcd}{}_a = \alpha_b \big( X^{J_1 I_2 I_3}_c X^{I_1 J_2 J_3}_d + X^{I_1 J_1 I_3}_c X^{I_2 J_2 J_3}_d + X^{I_1 I_2 J_1}_c X^{I_3 J_2 J_3}_d \big) \, d^{bcd}{}_a \, , \label{Useful_Id_5}
\end{equation}
\begin{equation}
\alpha_e \beta_f X^{J_1 J_2 J_3}_g X^I_c \, d^{efg}{}_d f^{cdb}{}_a = \alpha_e \beta_f \big( X^{I J_2 J_3}_g X^{J_1}_c + X^{J_1 I J_3}_g X^{J_2}_c + X^{J_1 J_2 I}_g X^{J_3}_c \big) \, d^{efg}{}_d f^{cdb}{}_a \, , \label{Useful_Id_4}
\end{equation}
where $\alpha$ and $\beta$ are either $\psi$, $DX$ or $[X,X,X]$. Using these identities it is possible to show that the additional terms are either identically zero or proportional to terms we have already listed.

\subsection{\sl Invariance of the Lagrangian} \label{Invariance}

We want to determine the coefficients for which the BLG Lagrangian together with its $1/T_{M2}$ correction given in \eref{L_Higher} is maximally supersymmetric. As the BLG Lagrangian is invariant under the lowest order supersymmetries i.e.\ $\delta \mathcal{L}_{BLG}=0$, the full corrected Lagrangian varies into 
\begin{equation}
\tilde{\delta} \mathcal{L} = \delta' \mathcal{L}_{BLG} + \delta \mathcal{L}_{1/T_{M2}} + \mathcal{O} \left( \sfrac{1}{(T_{M2})^2} \right) = \tilde{\delta} \mathcal{L}_4 + \tilde{\delta} \mathcal{L}_3 + \tilde{\delta} \mathcal{L}_2 + \tilde{\delta} \mathcal{L}_1 + \tilde{\delta} \mathcal{L}_0 \, ,
\end{equation}
where we ignore $\mathcal{O} \left( 1/(T_{M2})^2 \right)$ terms. The subscript in $\tilde{\delta} \mathcal{L}_n$ enumerates the total number of covariant derivatives acting on the fields and because the terms in $\tilde{\delta} \mathcal{L}_n$ are independent of those in any other $\tilde{\delta} \mathcal{L}_m$, invariance of the full Lagrangian means each $\tilde{\delta} \mathcal{L}_n$ must be invariant up to total derivatives.\footnote{There is the possibility that $\tilde{\delta} \mathcal{L}=0$ only after terms are removed using $1/T_{M2} \ \times$ lowest order equations of motion (which are $\mathcal{O} ( 1/T^2_{M2} ))$, in which case the different $\tilde{\delta} \mathcal{L}_n$ are not independent. However, we find for the Euclidean theory that invariance does not require use of the lowest order field equations.}

When we insert the higher derivative supersymmetries which are of the form $\delta' \chi_a = \alpha_b \beta_c \gamma_d \, d^{bcd}{}_a$, into the varied kinetic terms in $\delta' \mathcal{L}_{BLG}$ we find Tr is promoted to STr because
\begin{align}
{\rm Tr} ( \phi \delta' \chi ) = \phi^a \delta' \chi_a = \phi^a \alpha_b \beta_c \gamma_d \, d^{bcd}{}_a = \phi_a \alpha_b \beta_c \gamma_d \, d^{abcd} = {\rm STr} \{ \phi \alpha \beta \gamma \} \, .
\end{align}
Inserting the higher derivative supersymmetries into the varied bosonic potential and Yukawa terms in $\delta' \mathcal{L}_{BLG}$ requires more manipulation:
\begin{align}
{\rm Tr} ( \phi [ \lambda , \varphi , \delta' \chi ] ) = \phi_g \lambda_e \varphi_f \delta' \chi_a f^{efag} = \phi_g \lambda_e \varphi_f \alpha_b \beta_c \gamma_d \, d^{bcd}{}_a f^{efag} = - \phi_g \lambda_e \varphi_f \alpha_b \beta_c \gamma_d \, d^{abcd} f^{efg}{}_a \, .
\end{align}
Using the gauge invariance condition in \eref{Gauge_Inv} we can write this as
\begin{align}
{\rm Tr} ( \phi [ \lambda , \varphi , \delta' \chi ] ) = {\rm STr} \{ \phi \beta \gamma [ \lambda , \varphi , \alpha ] + \phi \alpha \gamma [ \lambda , \varphi , \beta ] + \phi \alpha \beta [ \lambda , \varphi , \gamma ] \} =: {\rm STr} \{ \phi [ \lambda , \varphi , \alpha \beta \gamma ] \} \, .
\end{align}

We are now in a position where we can proceed to compute $\tilde{\delta} \mathcal{L}$. We start by investigating the terms in the variation of the full corrected Lagrangian which contain four covariant derivatives. These come from
\begin{align}
\nonumber
\tilde{\delta} \mathcal{L}_{4} =\sfrac{1}{T_{M2}} {\rm STr} \Big\{ &- D_\mu ( \delta'_{2 DX} X^I ) D^\mu X^I + \sfrac{i}{2} \delta'_{3 DX} \bar{\psi} \Gamma^\mu D_\mu \psi + \sfrac{i}{2} \bar{\psi} \Gamma^\mu D_\mu ( \delta'_{3 DX} \psi ) + \sfrac{1}{2} \varepsilon^{\mu\nu\lambda} F_{\nu \lambda} \delta'_{2DX} \tilde{A}_\mu \\
\nonumber
&+ 4 \bold{a} \, D^\mu ( \delta X^I ) D_\mu X^J  D^\nu X^J D_\nu X^I + 4 \bold{b} \, D^\mu ( \delta X^I ) D_\mu X^I D^\nu X^J D_\nu X^J \\
\nonumber
&+ i \hat{\bold{d}} \, \delta_{1 DX} \bar{\psi} \Gamma_\mu \Gamma^{IJ} D_\nu\psi {D}^\mu X^{I} {D}^\nu X^J + i \hat{\bold{d}} \, \bar{\psi} \Gamma_\mu \Gamma^{IJ} D_\nu ( \delta_{1 DX}  \psi ) {D}^\mu X^{I} {D}^\nu X^J \\
&+ i \hat{\bold{e}} \, \delta_{1 DX}  \bar{\psi}\Gamma_\mu D^\nu\psi {D}^\mu X^{I} {D}_\nu X^I + i \hat{\bold{e}} \, \bar{\psi}\Gamma_\mu D^\nu ( \delta_{1 DX}  \psi ) {D}^\mu X^{I} {D}_\nu X^I \Big\} \, .
\end{align}
Note that the gauge field strength contributes two derivatives through its definition as the commutator of covariant derivatives. We have also split the lowest order fermion supersymmetry into $\delta \psi = \delta_{0DX} \psi + \delta_{1DX} \psi$ with $\delta_{0DX} \psi = \Gamma^\mu \Gamma^I \epsilon \, D_\mu X^I$ and $\delta_{0DX} \psi = - \sfrac{1}{6} \Gamma^{IJK} \epsilon \, X^{IJK}$. The next steps in the calculation are to insert the appropriate supersymmetry transformations, canonically reorder $\bar{\psi}$ and $\epsilon$ using the spinor flip condition \eref{Spinor_Flip} and then commute the worldvolume $\Gamma$-matrices through the transverse ones. Doing all this gives
\begin{align}
\nonumber
\tilde{\delta} \mathcal{L}_{4} = \sfrac{1}{T_{M2}} {\rm STr} \Big\{ &- i s_1 \bar{\epsilon} \Gamma^{IJK} \Gamma^{\lambda \mu} D_\nu ( \psi D_\lambda X^J D_\mu X^K ) D^\nu X^I - i s_2 \bar{\epsilon} \Gamma^I \Gamma^{\lambda \mu} D_\nu ( \psi D_\lambda X^J D_\mu X^I ) D^\nu X^J \\
\nonumber
&- i s_3 \bar{\epsilon} \Gamma^I D_\mu ( \psi D_\nu X^J D^\nu X^I ) D^\mu X^J - i s_4 \bar{\epsilon} \Gamma^I D_\mu ( \psi D_\nu X^J D^\nu X^J ) D^\mu X^I \\
\nonumber
&- \sfrac{i}{2} f_1 \bar{\epsilon} \Gamma^{IJK} \Gamma^{\nu \lambda \rho} \Gamma^\mu D_\mu \psi D_\nu X^I D_\lambda X^J D_\rho X^K + \sfrac{i}{2} f_1 \bar{\epsilon} \Gamma^{IJK} \Gamma^{\nu \lambda \rho} \Gamma^\mu \psi D_\mu ( D_\nu X^I D_\lambda X^J D_\rho X^K ) \\
\nonumber
&- \sfrac{i}{2} f_2 \bar{\epsilon} \Gamma^I \Gamma^\lambda \Gamma^\mu D_\mu \psi D_\lambda X^J D_\nu X^J D^\nu X^I + \sfrac{i}{2} f_2 \bar{\epsilon} \Gamma^I \Gamma^\lambda \Gamma^\mu \psi D_\mu ( D_\lambda X^J D_\nu X^J D^\nu X^I ) \\
\nonumber
&- \sfrac{i}{2} f_3 \bar{\epsilon} \Gamma^I \Gamma^\lambda \Gamma^\mu D_\mu \psi D_\lambda X^I D_\nu X^J D^\nu X^J + \sfrac{i}{2} f_3 \bar{\epsilon} \Gamma^I \Gamma^\lambda \Gamma^\mu \psi D_\mu ( D_\lambda X^I D_\nu X^J D^\nu X^J ) \\
\nonumber
&- \sfrac{i}{2} g_1 \varepsilon^{\mu \rho \sigma} \bar{\epsilon} \Gamma^I \Gamma_\mu \psi D_\nu X^J D^\nu X^J ( \tilde{F}_{\rho \sigma} X^I ) - \sfrac{i}{2} g_2 \varepsilon^{\mu \rho \sigma} \bar{\epsilon} \Gamma^I \Gamma^\nu \psi D_\mu X^J D_\nu X^J ( \tilde{F}_{\rho \sigma} X^I ) \\
\nonumber
&- \sfrac{i}{2} g_3 \varepsilon^{\mu \rho \sigma} \bar{\epsilon} \Gamma^J \Gamma^\nu \psi D_\mu X^J D_\nu X^I ( \tilde{F}_{\rho \sigma} X^I ) - \sfrac{i}{2} g_4 \varepsilon^{\mu \rho \sigma} \bar{\epsilon} \Gamma^J \Gamma^\nu \psi D_\mu X^I D_\nu X^J ( \tilde{F}_{\rho \sigma} X^I ) \\
\nonumber
&- \sfrac{i}{2} g_5 \varepsilon^{\mu \rho \sigma} \bar{\epsilon} \Gamma^J \Gamma_\mu \psi D_\nu X^J D^\nu X^I ( \tilde{F}_{\rho \sigma} X^I ) - \sfrac{i}{2} g_6 \varepsilon^{\mu \rho \sigma} \bar{\epsilon} \Gamma^J \Gamma_{\mu \nu \lambda} \psi D^\nu X^J D^\lambda X^I ( \tilde{F}_{\rho \sigma} X^I ) \\
\nonumber
&- \sfrac{i}{2} g_7 \varepsilon^{\mu \rho \sigma} \bar{\epsilon} \Gamma^{IJK} \Gamma_{\mu \nu \lambda} \psi D^\nu X^J D^\lambda X^K ( \tilde{F}_{\rho \sigma} X^I ) - \sfrac{i}{2} g_8 \varepsilon^{\mu \rho \sigma} \bar{\epsilon} \Gamma^{IJK} \Gamma^{\nu} \psi D_\mu X^J D_\nu X^K ( \tilde{F}_{\rho \sigma} X^I ) \\
\nonumber
&+ 4 i \bold{a} \, \bar{\epsilon} \Gamma^I D^\mu \psi D_\mu X^J  D^\nu X^J D_\nu X^I + 4 i \bold{b} \, \bar{\epsilon} \Gamma^I D^\mu \psi D_\mu X^I D^\nu X^J D_\nu X^J \\
\nonumber
&+ i \hat{\bold{d}} \, \bar{\epsilon} \Gamma^K \Gamma^{IJ} \Gamma^\lambda \Gamma^\mu D_\nu \psi D_\mu X^I D^\nu X^J D_\lambda X^K + i \hat{\bold{d}} \, \bar{\epsilon} \Gamma^K \Gamma^{IJ} \Gamma^\lambda \Gamma^\mu \psi D^\nu ( D_\lambda X^K ) D_\mu X^I D_\nu X^J \\
&+ i \hat{\bold{e}} \, \bar{\epsilon} \Gamma^J \Gamma^\lambda \Gamma^\mu D^\nu \psi D_\mu X^I D_\nu X^I D_\lambda X^J - i \hat{\bold{e}}  \bar{\epsilon} \Gamma^J \Gamma^\lambda \Gamma^\mu \psi D^\nu ( D_\lambda X^J ) D_\mu X^I D_\nu X^I \Big\}.
\end{align}
After using worldvolume $\Gamma$-matrix duality \eqref{wvol_duality_f} wherever $\varepsilon^{\mu \rho \sigma}$ occurs and then expanding out the $\Gamma$-matrices (this has been aided by use of Cadabra \cite{Peeters:2006kp}\cite{Peeters:2007wn}) we find the appearance of four distinct and independent types of $\Gamma$-matrix terms;  $\Gamma^{IJK} \Gamma^{\lambda \mu}$, $\Gamma^{IJK}$, $\Gamma^I \Gamma^{\lambda \mu}$ and $\Gamma^I$. We consider each of these types in turn.

We find the $\Gamma^{IJK} \Gamma^{\lambda \mu}$ terms to be
\begin{align}
\nonumber
\sfrac{1}{T_{M2}} {\rm STr} \Big\{ &+ i \left( - \sfrac{3}{2} f_1 - s_1 + \hat{\bold{d}} \right) \bar{\epsilon} \Gamma^{IJK} \Gamma^{\lambda \mu} D^\nu \psi D_\mu X^I D_\nu X^J D_\lambda X^K \\
\nonumber
&+ i \left( + \sfrac{3}{2} f_1 \right) \bar{\epsilon} \Gamma^{IJK} \Gamma^{\lambda \mu} \psi D^\nu ( D_\mu X^I D_\nu X^J D_\lambda X^K ) \\
\nonumber
&+ i \left(  - 2 s_1 +  \hat{\bold{d}} \right) \bar{\epsilon} \Gamma^{IJK} \Gamma^{\lambda \mu} \psi D^\nu ( D_\mu X^I ) D_\nu X^J D_\lambda X^K \\
&+ i \left( - g_8 \right) \bar{\epsilon} \Gamma^{IJK} \Gamma^{\lambda\mu} \psi D_\mu X^I D^\nu X^J ( \tilde{F}_{\nu \lambda} X^K ) \Big\} \, .
\end{align}
The first two lines combine to form a total derivative if they share the same coefficient. Hence we require $- \sfrac{3}{2} f_1 - s_1 + \hat{\bold{d}} = + \sfrac{3}{2} f_1$. The two remaining terms are invariant if $s_1 = + \sfrac{1}{2} \hat{\bold{d}}$ and $g_8=0$. The value for $s_1$ allows us to identify $f_1 = + \sfrac{1}{6} \hat{\bold{d}}$. 

The $\Gamma^{IJK}$ terms are
\begin{align}
\sfrac{1}{T_{M2}} {\rm STr} \Big\{ &- i \hat{\bold{d}} \, \bar{\epsilon} \Gamma^{IJK} \psi D^\mu ( D^\nu X^K ) D_\mu X^I D_\nu X^J + i g_7 \, \bar{\epsilon} \Gamma^{IJK} \psi D^\nu X^J D^\lambda X^K ( \tilde{F}_{\nu \lambda} X^I ) \Big\} \\
&= \sfrac{1}{T_{M2}} {\rm STr} \Big\{ + i ( - \hat{\bold{d}} + 2 g_7 ) \bar{\epsilon} \Gamma^{IJK} \psi D^\mu ( D^\nu X^K ) D_\mu X^I D_\nu X^J \Big\} \, ,
\end{align}
where we have made use of the definition $\tilde{F}_{\mu \nu} X = [ D_\mu , D_\nu ] X$ and relabelled dummy Lorentz indices. Invariance of the $\Gamma^{IJK}$ terms then follows if $g_7 = + \sfrac{1}{2} \hat{\bold{d}}$.

After some manipulation the $\Gamma^I \Gamma^{\lambda \mu}$ terms are
\begin{align}
\nonumber
\sfrac{1}{T_{M2}} {\rm STr} \Big\{ &+ i \left( - \sfrac{1}{2} f_2 \right) \bar{\epsilon} \Gamma^I \Gamma^{\lambda \mu} D_\mu \psi D_\lambda X^J D_\nu X^J D^\nu X^I \\
\nonumber
&+ i \left( - \sfrac{1}{2} f_2 \right) \bar{\epsilon} \Gamma^I \Gamma^{\lambda \mu} \psi D_\mu ( D_\lambda X^J ) D_\nu X^J D^\nu X^I \\
\nonumber
&+ i \left( + \sfrac{1}{2} f_2 - \hat{\bold{d}} \right) \bar{\epsilon} \Gamma^I \Gamma^{\lambda \mu} \psi D_\lambda X^J D_\mu ( D_\nu X^J ) D^\nu X^I \\
\nonumber
&+ i \left( + \sfrac{1}{2} f_2 + \hat{\bold{e}} - s_2 \right) \bar{\epsilon} \Gamma^I \Gamma^{\lambda \mu} \psi D_\lambda X^J D_\nu X^J D_\mu ( D^\nu X^I ) \\[8pt]
\nonumber
&+ i \left( - \sfrac{1}{2} f_3 \right) \bar{\epsilon} \Gamma^I \Gamma^{\lambda \mu} D_\mu \psi D_\lambda X^I D_\nu X^J D^\nu X^J \\
\nonumber
&+ i \left( - \sfrac{1}{2} f_3 \right) \bar{\epsilon} \Gamma^I \Gamma^{\lambda \mu} \psi D_\mu ( D_\lambda X^I ) D_\nu X^J D^\nu X^J \\
\nonumber
&+ i \left( + \sfrac{1}{2} f_3 + \sfrac{1}{2} \hat{\bold{d}} + \sfrac{1}{2} s_2 \right) \bar{\epsilon} \Gamma^I \Gamma^{\lambda \mu} \psi D_\lambda X^I D_\mu ( D_\nu X^J D^\nu X^J ) \\[8pt]
\nonumber
&+ i \left( - \hat{\bold{d}} - \hat{\bold{e}} - s_2 \right) \bar{\epsilon} \Gamma^I \Gamma^{\lambda \mu} D^\nu \psi D_\mu X^I D_\nu X^J D_\lambda X^J \\
\nonumber
&+ i \left( + \sfrac{1}{2} f_2 + \sfrac{1}{2} g_3 + \sfrac{1}{2} g_4 + \sfrac{1}{2} g_5 \right) \bar{\epsilon} \Gamma^I \Gamma^{\lambda \mu} \psi ( \tilde{F}_{\mu \lambda} X^J ) D^\nu X^I D_\nu X^J \\
\nonumber
&+ i \left( + \sfrac{1}{2} f_3 + \sfrac{1}{2} g_1 + \sfrac{1}{2} g_2 \right) \bar{\epsilon} \Gamma^I \Gamma^{\lambda \mu} \psi ( \tilde{F}_{\mu \lambda} X^I ) D_\nu X^J D^\nu X^J \\
\nonumber
&+ i \left( - \hat{\bold{d}} - s_2 - g_3 \right) \bar{\epsilon} \Gamma^I \Gamma^{\lambda \mu} \psi ( \tilde{F}_{\mu \nu} X^J ) D_\lambda X^I D^\nu X^J \\
\nonumber
&+ i \left( - \hat{\bold{e}} + s_2 - g_2 \right) \bar{\epsilon} \Gamma^I \Gamma^{\lambda \mu} \psi ( \tilde{F}_{\mu \nu} X^I ) D_\lambda X^J D^\nu X^J \\
&+ i \left( + \hat{\bold{d}} - g_4 \right) \bar{\epsilon} \Gamma^I \Gamma^{\lambda \mu} \psi ( \tilde{F}_{\mu \nu} X^J ) D^\nu X^I D_\lambda X^J \Big\} \, .
\end{align}
The first seven lines can be written as two distinct total derivatives provided
\begin{align}
&- \sfrac{1}{2} f_2 = + \sfrac{1}{2} f_2 - \hat{\bold{d}} = + \sfrac{1}{2} f_2 + \hat{\bold{e}} - s_2 \, , \\
&- \sfrac{1}{2} f_3 = + \sfrac{1}{2} f_3 + \sfrac{1}{2} \hat{\bold{d}} + \sfrac{1}{2} s_2 \, .
\end{align}
The remaining terms vanish if 
\begin{align}
0=&- \hat{\bold{d}} - \hat{\bold{e}} - s_2 \, , \\
0=&+ \sfrac{1}{2} f_2 + \sfrac{1}{2} g_3 + \sfrac{1}{2} g_4 + \sfrac{1}{2} g_5 \, , \\
0=&+ \sfrac{1}{2} f_3 + \sfrac{1}{2} g_1 + \sfrac{1}{2} g_2 \, , \\
0=&- \hat{\bold{d}} - s_2 - g_3 \, , \\
0=&- \hat{\bold{e}} + s_2 - g_2 \, , \\
0=&+ \hat{\bold{d}} - g_4 \, .
\end{align}
The solution to these simultaneous equations is
\begin{equation}
f_2 = + \hat{\bold{d}} \, , \quad f_3 = - \sfrac{1}{2} \hat{\bold{d}} \, , \quad s_2 = 0 \, , \quad \hat{\bold{e}} = - \hat{\bold{d}} \ , \label{Intermediate_1}
\end{equation}
\begin{equation}
g_1 = - \sfrac{1}{2} \hat{\bold{d}} \, , \quad g_2 = + \hat{\bold{d}} \, , \quad g_3 = - \hat{\bold{d}} \, , \quad g_4 = + \hat{\bold{d}} \, , \quad g_5 = - \hat{\bold{d}} \, .
\end{equation}
Finally, the $\Gamma^I$ terms can be manipulated to arrive at
\begin{align}
\nonumber
\sfrac{1}{T_{M2}} {\rm STr} \Big\{ &+ i \left( - \sfrac{1}{2} f_2 - s_3 + 4 \bold{a} - \hat{\bold{d}} + \hat{\bold{e}} \right) \bar{\epsilon} \Gamma^I D^\mu \psi D^\nu X^I D_\mu X^J D_\nu X^J \\
\nonumber
&+ i \left( + \sfrac{1}{2} f_2 - s_3 - \hat{\bold{e}} \right) \bar{\epsilon} \Gamma^I \psi D^\mu ( D^\nu X^I ) D_\mu X^J D_\nu X^J \\
\nonumber
&+ i \left( + \sfrac{1}{2} f_2 \right) \bar{\epsilon} \Gamma^I \psi D^\nu X^I D^\mu ( D_\mu X^J ) D_\nu X^J \\
\nonumber
&+ i \left( + \sfrac{1}{2} f_2 - s_3 - g_6 - \hat{\bold{d}} \right) \bar{\epsilon} \Gamma^I \psi D^\nu X^I D_\mu X^J D^\mu ( D_\nu X^J ) \\[8pt]
\nonumber
&+i \left( - \sfrac{1}{2} f_3 - s_4 + 4 \bold{b} + \hat{\bold{d}} \right) \bar{\epsilon} \Gamma^I D^\mu \psi D_\mu X^I D_\nu X^J D^\nu X^J \\
\nonumber
&+ i \left( + \sfrac{1}{2} f_3 \right) \bar{\epsilon} \Gamma^I \psi D^\mu ( D_\mu X^I ) D_\nu X^J D^\nu X^J \\
&+ i \left( + \sfrac{1}{2} f_3 - s_4 + \sfrac{1}{2} g_6 + \sfrac{1}{2} \hat{\bold{d}} \right) \bar{\epsilon} \Gamma^I \psi D_\mu X^I D^\mu ( D_\nu X^J D^\nu X^J ) \Big\} \, . \label{4DX_Gamma_I}
\end{align}
We see that the first four lines combine to form a total derivative if
\begin{equation}
- \sfrac{1}{2} f_2 - s_3 + 4 \bold{a} - \hat{\bold{d}} + \hat{\bold{e}} = + \sfrac{1}{2} f_2 - s_3 - \hat{\bold{e}} = + \sfrac{1}{2} f_2 = + \sfrac{1}{2} f_2 - s_3 - g_6 - \hat{\bold{d}} \, .
\end{equation}
The last three lines form another total derivative provided
\begin{equation}
- \sfrac{1}{2} f_3 - s_4 + 4 \bold{b} + \hat{\bold{d}} = + \sfrac{1}{2} f_3 = + \sfrac{1}{2} f_3 - s_4 + \sfrac{1}{2} g_6 + \sfrac{1}{2} \hat{\bold{d}} \, .
\end{equation}
Using the values for $f_2$, $f_3$ and $\hat{\bold{e}}$ in \eref{Intermediate_1} we can solve these latest simultaneous equations to discover
\begin{equation}
s_3 = + \hat{\bold{d}} \, , \quad s_4 = - \sfrac{1}{2} \hat{\bold{d}} \, , \quad g_6 = - 2 \hat{\bold{d}} \, , \quad \bold{a} = + \hat{\bold{d}} \ , \quad \bold{b} = - \sfrac{1}{2} \hat{\bold{d}} \, .
\end{equation}

To summarise, the four covariant derivative terms $\tilde{\mathcal{L}}_4$ are invariant up to boundary terms if the coefficients in the Lagrangian and supersymmetry transformations are given by
\begin{equation}
\bold{a} = + \hat{\bold{d}} \, , \quad \bold{b} = - \sfrac{1}{2} \hat{\bold{d}} \, , \quad \hat{\bold{e}} = - \hat{\bold{d}} \, , \label{4DX_L}
\end{equation}
\begin{equation}
f_1 = + \sfrac{1}{6} \hat{\bold{d}} \, , \quad f_2 = + \hat{\bold{d}} \, , \quad f_3 = - \sfrac{1}{2} \hat{\bold{d}} \, , \label{4DX_f}
\end{equation}
\begin{equation}
s_1 = + \sfrac{1}{2} \hat{\bold{d}} \, , \quad s_2 = 0 \, , \quad s_3 = + \hat{\bold{d}} \, , \quad s_4 = - \sfrac{1}{2} \hat{\bold{d}} \, , \label{4DX_s}
\end{equation}
\begin{equation}
\nonumber
g_1 = - \sfrac{1}{2} \hat{\bold{d}} \, , \quad g_2 = + \hat{\bold{d}} \, , \quad g_3 = - \hat{\bold{d}} \, , \quad g_4 = + \hat{\bold{d}} \, , 
\end{equation}
\begin{equation}
g_5 = - \hat{\bold{d}} \, , \quad g_6 = - 2 \hat{\bold{d}} \, , \quad g_7 = + \sfrac{1}{2} \hat{\bold{d}} \, , \quad g_8 = 0 \, . \label{4DX_g}
\end{equation}
The coefficients in Eqs.\,(\ref{4DX_f}) and (\ref{4DX_s}) satisfy the relations previously found by Low \cite{Low:2010ie}.

We now consider the terms in $\delta' \mathcal{L}_{BLG} + \delta \mathcal{L}_{1/T_{M2}}$ which contain a total of three covariant derivatives. These are,
\begin{align}
\nonumber
\tilde{\delta} \mathcal{L}_3 = \sfrac{1}{T_{M2}} {\rm STr} \Big\{ &- D_\mu ( \delta'_{1 DX} X^I ) D^\mu X^I + \delta'_{2 DX} \tilde{A}_\mu X^I D^\mu X^I \\
\nonumber
&+ \sfrac{i}{2} \delta'_{2 DX} \bar{\psi} \Gamma^\mu D_\mu \psi + \sfrac{i}{2} \bar{\psi} \Gamma^\mu D_\mu ( \delta'_{2 DX} \psi ) \\
\nonumber
&+ \sfrac{i}{4} \delta'_{3 DX} \bar{\psi} \Gamma_{IJ} [ X^I , X^J , \psi ] + \sfrac{i}{4} \bar{\psi} \Gamma_{IJ} [ X^I , X^J , \delta'_{3 DX} \psi ] \\
\nonumber
&+ \sfrac{1}{2} \varepsilon^{\mu \rho \sigma} F_{\rho \sigma} \delta'_{1DX} \tilde{A}_\mu \\
\nonumber
&-4 \bold{a} \, \delta \tilde{A}^\mu X^I D_\mu X^J  D^\nu X^J D_\nu X^I - 4 \bold{b} \, \delta \tilde{A}^\mu X^I D_\mu X^I D^\nu X^J D_\nu X^J \\
\nonumber
&+ 3 \bold{c} \, \varepsilon^{\mu\nu\lambda}\, [ \delta X^I , X^J , X^K ] D_\mu X^I D_\nu X^J D_\lambda X^K + 3 \bold{c} \, \varepsilon^{\mu\nu\lambda}\, X^{IJK} D_\mu ( \delta X^I ) D_\nu X^J D_\lambda X^K \\
\nonumber
&+ i \hat{\bold{d}} \, \delta_{0 DX} \bar{\psi} \Gamma_\mu \Gamma^{IJ} D_\nu\psi {D}^\mu X^{I} {D}^\nu X^J + i \hat{\bold{d}} \, \bar{\psi} \Gamma_\mu \Gamma^{IJ} D_\nu ( \delta_{0 DX} \psi ) {D}^\mu X^{I} {D}^\nu X^J \\
\nonumber
&+ i \hat{\bold{e}} \, \delta_{0 DX} \bar{\psi}\Gamma_\mu D^\nu\psi {D}^\mu X^{I} {D}_\nu X^I + i \hat{\bold{e}} \, \bar{\psi}\Gamma_\mu D^\nu ( \delta_{0 DX} \psi ) {D}^\mu X^{I} {D}_\nu X^I \\
\nonumber
&+ i \hat{\bold{f}} \, \delta_{1 DX} \bar{\psi}\Gamma^{IJKL} D_\nu\psi\; X^{IJK} {D}^\nu X^L + i \hat{\bold{f}} \, \bar{\psi}\Gamma^{IJKL} D_\nu ( \delta_{1 DX} \psi ) X^{IJK}  {D}^\nu X^L \\
\nonumber
&+ i \hat{\bold{g}} \, \delta_{1 DX} \bar{\psi}\Gamma^{IJ} D_\nu\psi\; X^{IJK}{D}^\nu X^K + i \hat{\bold{g}} \, \bar{\psi}\Gamma^{IJ} D_\nu ( \delta_{1 DX} \psi ) X^{IJK}{D}^\nu X^K \\
\nonumber
&+ i \hat{\bold{h}} \, \delta_{1 DX} \bar{\psi}\Gamma^{IJ}[X^J,X^{K},\psi] {D}^\mu X^{I} {D}_\mu X^K + i \hat{\bold{h}} \, \bar{\psi}\Gamma^{IJ}[X^J,X^{K}, \delta_{1 DX} \psi] {D}^\mu X^{I} {D}_\mu X^K \\
\nonumber
&+ i \hat{\bold{i}} \, \delta_{1 DX} \bar{\psi}\Gamma^{\mu\nu}[X^I,X^{J},\psi] {D}_\mu X^{I}{D}_\nu X^J + i \hat{\bold{i}} \, \bar{\psi}\Gamma^{\mu\nu}[X^I,X^{J}, \delta_{1 DX} \psi] {D}_\mu X^{I}{D}_\nu X^J \\
\nonumber
&+ i \hat{\bold{j}} \, \delta_{1 DX} \bar{\psi}\Gamma_{\mu\nu} \Gamma^{IJ} [X^J,X^{K},\psi] {D}^\mu X^{I}{D}^\nu X^K \\
&+ i \hat{\bold{j}} \, \bar{\psi}\Gamma_{\mu\nu} \Gamma^{IJ} [X^J,X^{K},\delta_{1 DX} \psi] {D}^\mu X^{I}{D}^\nu X^K \Big\} \, .
\end{align}
Once again we insert the appropriate supersymmetry transformations, canonically reorder $\bar{\psi}$ and $\epsilon$ using the spinor flip condition \eref{Spinor_Flip} and then commute the worldvolume $\Gamma$-matrices through the transverse ones. The result is
\begin{align}
\nonumber
\sfrac{1}{T_{M2}} {\rm STr} \Big\{ &- i s_5 \bar{\epsilon} \Gamma^{IJKLM} \Gamma^\mu D^\nu ( \psi D_\mu X^J X^{KLM} ) D_\nu X^I - i s_6 \bar{\epsilon} \Gamma^{KLM} \Gamma^\mu D^\nu ( \psi D_\mu X^I X^{KLM} ) D_\nu X^I \\
\nonumber
&- i s_7 \bar{\epsilon} \Gamma^{JLM} \Gamma^\mu D^\nu ( \psi D_\mu X^J X^{ILM} ) D_\nu X^I - i s_8 \bar{\epsilon} \Gamma^{ILM} \Gamma^\mu D^\nu ( \psi D_\mu X^J X^{JLM} ) D_\nu X^I \\
\nonumber
&- i s_9 \bar{\epsilon} \Gamma^{M} \Gamma^\mu D^\nu ( \psi D_\mu X^J X^{IJM} ) D_\nu X^I \\
\nonumber
&- i g_1 \bar{\epsilon} \Gamma^J \Gamma^\mu \psi [ X^J ,  X^I , D^\mu X^I ] D_\nu X^K D^\nu X^K - i g_2 \bar{\epsilon} \Gamma^J \Gamma^\nu \psi [ X^J , X^I , D^\mu X^I ] D_\mu X^K D_\nu X^K \\
\nonumber
&- i g_3 \bar{\epsilon} \Gamma^J \Gamma^\nu \psi [ X^K , X^I , D^\mu X^I ] D_\mu X^J D_\nu X^K - i g_4 \bar{\epsilon} \Gamma^J \Gamma^\nu \psi [ X^K , X^I , D^\mu X^I ] D_\mu X^K D_\nu X^J \\
\nonumber
&- i g_5 \bar{\epsilon} \Gamma^J \Gamma^\mu \psi [ X^K , X^I , D_\mu X^I ] D_\nu X^J D^\nu X^K - i g_6 \bar{\epsilon} \Gamma^J \Gamma^{\mu \nu \lambda} \psi [ X^K , X^I , D_\mu X^I ] D_\nu X^J D_\lambda X^K \\
\nonumber
&- i g_7 \bar{\epsilon} \Gamma^{JKL} \Gamma^{\mu \nu \lambda} \psi [ X^J , X^I , D_\mu X^I ] D_\nu X^K D_\lambda X^L - i g_8 \bar{\epsilon} \Gamma^{JKL} \Gamma^\nu \psi [ X^J , X^I , D^\mu X^I ] D_\mu X^K D_\nu X^L \\
\nonumber
&+ \sfrac{i}{2} f_4 \bar{\epsilon} \Gamma^{IJKLM} \Gamma^{\nu \lambda} \Gamma^\mu D_\mu \psi D_\nu X^I D_\lambda X^J X^{KLM} - \sfrac{i}{2} f_4 \bar{\epsilon} \Gamma^{IJKLM} \Gamma^{\nu \lambda} \Gamma^\mu \psi D_\mu ( D_\nu X^I D_\lambda X^J X^{KLM} ) \\
\nonumber
&- \sfrac{i}{2} f_5 \bar{\epsilon} \Gamma^{KLM} \Gamma^{\nu \lambda} \Gamma^\mu D_\mu \psi D_\nu X^J D_\lambda X^K X^{JLM} + \sfrac{i}{2} f_5 \bar{\epsilon} \Gamma^{KLM} \Gamma^{\nu \lambda} \Gamma^\mu \psi D_\mu ( D_\nu X^J D_\lambda X^K X^{JLM} ) \\
\nonumber
&+ \sfrac{i}{2} f_6 \bar{\epsilon} \Gamma^M \Gamma^{\nu \lambda} \Gamma^\mu D_\mu \psi D_\nu X^J D_\lambda X^K X^{JKM} - \sfrac{i}{2} f_6 \bar{\epsilon} \Gamma^M \Gamma^{\nu \lambda} \Gamma^\mu \psi D_\mu ( D_\nu X^J D_\lambda X^K X^{JKM} ) \\
\nonumber
&+ \sfrac{i}{2} f_7 \bar{\epsilon} \Gamma^{KLM} \Gamma^\mu D_\mu \psi D_\nu X^J D^\nu X^J X^{KLM} - \sfrac{i}{2} f_7 \bar{\epsilon} \Gamma^{KLM} \Gamma^\mu \psi D_\mu ( D_\nu X^J D^\nu X^J X^{KLM} ) \\
\nonumber
&+ \sfrac{i}{2} f_8 \bar{\epsilon} \Gamma^{KLM} \Gamma^\mu D_\mu \psi D_\nu X^J D^\nu X^K X^{JLM} - \sfrac{i}{2} f_8 \bar{\epsilon} \Gamma^{KLM} \Gamma^\mu \psi D_\mu ( D_\nu X^J D^\nu X^K X^{JLM} ) \\
\nonumber
&- \sfrac{i}{4} f_1 \bar{\epsilon} \Gamma^{KLM} \Gamma^{IJ} \Gamma^{\mu \nu \lambda} [ X^I , X^J , \psi ] D_\mu X^K D_\nu X^L D_\lambda X^M \\
\nonumber
&+ \sfrac{i}{4} f_1 \bar{\epsilon} \Gamma^{KLM} \Gamma^{IJ} \Gamma^{\mu \nu \lambda} \psi [ X^I , X^J , D_\mu X^K D_\nu X^L D_\lambda X^M ] \\
\nonumber
&- \sfrac{i}{4} f_2 \bar{\epsilon} \Gamma^K \Gamma^{IJ} \Gamma^\mu [ X^I , X^J , \psi ] D_\mu X^L D_\nu X^L D^\nu X^K + \sfrac{i}{4} f_2 \bar{\epsilon} \Gamma^K \Gamma^{IJ} \Gamma^\mu \psi [ X^I , X^J , D_\mu X^L D_\nu X^L D^\nu X^K ] \\
\nonumber
&- \sfrac{i}{4} f_3 \bar{\epsilon} \Gamma^K \Gamma^{IJ} \Gamma^\mu [ X^I , X^J , \psi ] D_\mu X^K D_\nu X^L D^\nu X^L + \sfrac{i}{4} f_3 \bar{\epsilon} \Gamma^K \Gamma^{IJ} \Gamma^\mu \psi [ X^I , X^J , D_\mu X^K D_\nu X^L D^\nu X^L ] \\
\nonumber
&+ \sfrac{i}{2} g_9 \varepsilon^{\mu \rho \sigma} \bar{\epsilon} \Gamma^{KLM} \Gamma_{\mu \nu} \psi D^\nu X^I X^{KLM} ( \tilde{F}_{\rho \sigma} X^I ) + \sfrac{i}{2} g_{10} \varepsilon^{\mu \rho \sigma} \bar{\epsilon} \Gamma^{JLM} \Gamma_{\mu \nu} \psi D^\nu X^J X^{ILM} ( \tilde{F}_{\rho \sigma} X^I ) \\
\nonumber
&+ \sfrac{i}{2} g_{11} \varepsilon^{\mu \rho \sigma} \bar{\epsilon} \Gamma^{M} \Gamma_{\mu \nu} \psi D^\nu X^J X^{IJM} ( \tilde{F}_{\rho \sigma} X^I ) + \sfrac{i}{2} g_{12} \varepsilon^{\mu \rho \sigma} \bar{\epsilon} \Gamma^{KLM} \psi D_\mu X^I X^{KLM} ( \tilde{F}_{\rho \sigma} X^I ) \\	
\nonumber
&+ \sfrac{i}{2} g_{13} \varepsilon^{\mu \rho \sigma} \bar{\epsilon} \Gamma^{JLM} \psi D_\mu X^J X^{ILM} ( \tilde{F}_{\rho \sigma} X^I ) + \sfrac{i}{2} g_{14} \varepsilon^{\mu \rho \sigma} \bar{\epsilon} \Gamma^{M} \psi D_\mu X^J X^{IJM} ( \tilde{F}_{\rho \sigma} X^I ) \\
\nonumber
&- 4 i \bold{a} \, \bar{\epsilon} \Gamma^K \Gamma^\mu [X^K, X^I ,\psi ] D_\mu X^J D^\nu X^J D_\nu X^I - 4 i \bold{b} \, \bar{\epsilon} \Gamma^K \Gamma^\mu [X^K, X^I ,\psi ] D_\mu X^I D^\nu X^J D_\nu X^J \\
\nonumber
&+ 3i \bold{c} \, \varepsilon^{\mu\nu\lambda}\, \bar{\epsilon} \Gamma^I [ \psi , X^J , X^K ] D_\mu X^I D_\nu X^J D_\lambda X^K + 3i \bold{c} \, \varepsilon^{\mu\nu\lambda}\, \bar{\epsilon} \Gamma^I D_\mu \psi D_\nu X^J D_\lambda X^K X^{IJK} \\
\nonumber
&- \sfrac{i \hat{\bold{d}}}{6} \, \bar{\epsilon} \Gamma^{KLM} \Gamma^{IJ} \Gamma^\mu D_\nu \psi D_\mu X^I D^\nu X^J X^{KLM} - \sfrac{i \hat{\bold{d}}}{6} \, \bar{\epsilon} \Gamma^{KLM} \Gamma^{IJ} \Gamma^\mu \psi D_\nu ( X^{KLM} ) D_\mu X^I D^\nu X^J \\
\nonumber
&- \sfrac{i \hat{\bold{e}}}{6} \, \bar{\epsilon} \Gamma^{KLM} \Gamma^\mu D^\nu \psi D_\mu X^I D_\nu X^I X^{KLM} + \sfrac{i \hat{\bold{e}}}{6} \, \bar{\epsilon} \Gamma^{KLM} \Gamma^\mu \psi D^\nu ( X^{KLM}) D_\mu X^I D_\nu X^I \\
\nonumber
&+ i \hat{\bold{f}} \, \bar{\epsilon} \Gamma^M \Gamma^{IJKL} \Gamma^\mu D_\nu \psi X^{IJK} D^\nu X^L D_\mu X^M + i \hat{\bold{f}} \, \bar{\epsilon} \Gamma^M \Gamma^{IJKL} \Gamma^\mu \psi D_\nu ( D_\mu X^M ) X^{IJK} D^\nu X^L \\
\nonumber
&+ i \hat{\bold{g}} \, \bar{\epsilon} \Gamma^L \Gamma^{IJ} \Gamma^\mu D_\nu \psi X^{IJK} D^\nu X^K D_\mu X^L - i \hat{\bold{g}} \, \bar{\epsilon} \Gamma^L \Gamma^{IJ} \Gamma^\mu \psi D_\nu ( D_\mu X^L ) X^{IJK} D^\nu X^K \\
\nonumber
&+ i \hat{\bold{h}} \, \bar{\epsilon} \Gamma^L \Gamma^{IJ} \Gamma^\mu [X^J,X^{K},\psi] D^\nu X^I D_\nu X^K D_\mu X^L - i \hat{\bold{h}} \, \bar{\epsilon} \Gamma^L \Gamma^{IJ} \Gamma^\mu \psi [X^J,X^K, D_\mu X^L ] D^\nu X^I D_\nu X^K \\
\nonumber
&+ i \hat{\bold{i}} \, \bar{\epsilon} \Gamma^K \Gamma^\lambda \Gamma^{\mu\nu} [X^I,X^J,\psi] D_\mu X^ID_\nu X^J D_\lambda X^K - i \hat{\bold{i}} \, \bar{\epsilon} \Gamma^K \Gamma^\lambda \Gamma^{\mu\nu} \psi [X^I,X^J, D_\lambda X^K ] D_\mu X^I D_\nu X^J \\
\nonumber
&+ i \hat{\bold{j}} \, \bar{\epsilon} \Gamma^L \Gamma^{IJ} \Gamma^\lambda \Gamma^{\mu\nu} [X^J,X^{K},\psi] D_\mu X^I D_\nu X^K D_\lambda X^L \\
&+ i \hat{\bold{j}} \, \bar{\epsilon} \Gamma^L \Gamma^{IJ} \Gamma^\lambda \Gamma^{\mu\nu} \psi [X^J,X^K,D_\lambda X^L ] D_\mu X^I D_\nu X^K \Big\} \, .
\end{align}
We have omitted the calculations due to their length however, after using worldvolume dualisation and performing the $\Gamma$-matrix algebra to we find all the terms in $\tilde{\delta} \mathcal{L}_3$ can be assembled into total derivatives or made to vanish through the gauge invariance condition in \eref{Gauge_Inv}. As in $\tilde{\delta} \mathcal{L}_4$, this requires the coefficients to satisfy certain constraints. Using the coefficient data from $\tilde{\delta} \mathcal{L}_4$ we can solve these additional simultaneous equations to find that $\tilde{\delta} \mathcal{L}_3$ is invariant if
\begin{equation}
\bold{c} = + \sfrac{2}{3} \hat{\bold{d}} \, , \quad \hat{\bold{f}} = + \sfrac{1}{6} \hat{\bold{d}} \, , \quad \hat{\bold{g}} = - \sfrac{1}{2} \hat{\bold{d}} \, , \quad \hat{\bold{h}} = + \hat{\bold{d}} \, , \quad \hat{\bold{i}} = - \hat{\bold{d}} \, , \quad \hat{\bold{j}} = - \hat{\bold{d}} \, , \label{3DX_L}
\end{equation}
\begin{equation}
f_4 = + \sfrac{1}{12} \hat{\bold{d}} \, , \quad f_5 = 0 \, , \quad f_6 = - \sfrac{3}{2} \hat{\bold{d}} \, , \quad f_7 = - \sfrac{1}{12} \hat{\bold{d}} \, , \quad f_8 = + \sfrac{1}{2} \hat{\bold{d}} \, , \label{3DX_f}
\end{equation}
\begin{equation}
s_5 = + \sfrac{1}{6} \hat{\bold{d}} \, , \quad s_6 = 0 \, , \quad s_7 = 0 \, , \quad s_8 = 0 \, , \quad s_9 = + \hat{\bold{d}} \, , \label{3DX_s}
\end{equation}
\begin{equation}
g_9 = 0 \, , \quad g_{10} = + \sfrac{1}{2} \hat{\bold{d}} \, , \quad g_{11} = 0 \, , \quad g_{12} = 0 \, , \quad g_{13} = 0 \, , \quad g_{14} = + \hat{\bold{d}} \, . \label{3DX_g}
\end{equation}

Demonstrating invariance of the terms $\tilde{\delta} \mathcal{L}_2$, $\tilde{\delta} \mathcal{L}_1$ and $\tilde{\delta} \mathcal{L}_0$ proceeds analogously to $\tilde{\delta} \mathcal{L}_4$ and $\tilde{\delta} \mathcal{L}_3$ only now the presence of two or more 3-brackets means we can manipulate terms using the fundamental identity \eref{FI} as well as the $\mathcal{A}_4$ identities in Eqs.\,(\ref{Useful_Id}) and (\ref{Useful_Id2}). We find invariance of $\tilde{\delta} \mathcal{L}_2$ is achieved if,
\begin{equation}
\bold{d} = + \hat{\bold{d}} \, , \quad \bold{e} = - \sfrac{1}{6} \hat{\bold{d}} \, , \quad \hat{\bold{k}} = + \sfrac{1}{2} \hat{\bold{d}} \, , \quad \hat{\bold{l}} = - \sfrac{1}{2} \hat{\bold{d}} \, , \quad \hat{\bold{n}} = - \sfrac{1}{2} \hat{\bold{d}} \, , \label{2DX_L}
\end{equation}
\begin{equation}
f_9 = - \sfrac{1}{12} \hat{\bold{d}} \, , \quad f_{10} = + \sfrac{1}{2} \hat{\bold{d}} \, , \label{2DX_f}
\end{equation}
\begin{equation}
s_{10} = - \sfrac{1}{12} \hat{\bold{d}} \, , \quad s_{11} = + \sfrac{1}{2} \hat{\bold{d}} \, , \label{2DX_s}
\end{equation}
\begin{equation}
g_{15} = - \sfrac{1}{12} \hat{\bold{d}} \, . \label{2DX_g}
\end{equation}

The additional constraints from invariance of the $\tilde{\delta} \mathcal{L}_1$ terms are $\hat{\bold{p}} = - \sfrac{1}{4} \hat{\bold{d}}$ and $f_{11} = + \sfrac{1}{72} \hat{\bold{d}}$ whilst the $\tilde{\delta} \mathcal{L}_0$ terms require $\bold{f} = + \sfrac{1}{72} \hat{\bold{d}}$.

We have been able to determine all the arbitrary coefficients in the order $1/T_{M2}$ Lagrangian and supersymmetry transformations up to a scale factor parametrised by $\hat{\bold{d}}$. The numerical value for $\hat{\bold{d}}$ can be fixed by reference to the action for a single M2-brane in \eref{M2_action}. We have seen in moving from a single M2-brane to multiple M2-branes the lowest order scalar kinetic terms are generalised as 
\begin{equation}
- \sfrac{1}{2} \partial_\mu X^I \partial^\mu X^I \rightarrow{\rm Tr} \big( - \sfrac{1}{2} D_\mu X^I D^\mu X^I \big) \, .
\end{equation}
It seems reasonable that the $1/T_{M2}$ corrections in \eref{M2_action} have a similar generalisation so that
\begin{align}
\nonumber
\sfrac{1}{T_{M2}} \Big( + \sfrac{1}{4} \partial_\mu X^I \partial^\mu X^J & \partial_\nu X^I \partial^\nu X^J - \sfrac{1}{8} \partial_\mu X^I \partial^\mu X^I \partial_\nu X^J \partial^\nu X^J \Big) \\
\rightarrow& \sfrac{1}{T_{M2}} {\rm STr} \Big( + \sfrac{1}{4} D_\mu X^I D^\mu X^J D_\nu X^I D^\nu X^J - \sfrac{1}{8} D_\mu X^I D^\mu X^I D_\nu X^J D^\nu X^J \Big) \, .
\end{align}
Thus, comparing with the $1/T_{M2}$ ansatz in \eref{L_Higher} we find $\bold{a} = + \hat{\bold{d}} = + \sfrac{1}{4}$ and $\bold{b} = -\sfrac{1}{2} \hat{\bold{d}} = - \sfrac{1}{8}$ which implies $\hat{\bold{d}} = + \sfrac{1}{4}$. Having fixed the scale parameter the remaining numerical values of the coefficients are\footnote{In comparing our results for the Lagrangian coefficients to those of \cite{Ezhuthachan:2009sr} we find some differences: although the values for the coefficients $\bold{a}$, $\bold{b}$, $\bold{d}$-$\bold{f}$ and $\hat{\bold{d}}$-$\hat{\bold{g}}$ match, in \cite{Ezhuthachan:2009sr} non-zero values are assigned to $\hat{\bold{m}}$ and $\hat{\bold{o}}$ whereas we have found they should be dropped from the $\mathcal{A}_4$ Lagrangian. For the remaining coefficients, $\bold{c}$ and $\hat{\bold{h}}$-$\hat{\bold{p}}$, we find the absolute values match but that there is disagreement over signs.},
\begin{align}
\nonumber
s_1 &= + \sfrac{1}{8} \, , & &f_1 = + \sfrac{1}{24} \, , & &g_1 = - \sfrac{1}{8} \, , & &\bold{a} = + \sfrac{1}{4} \, , \\
\nonumber
 s_2 &=  0 \, , & &f_2 = + \sfrac{1}{4} \, , & &g_2 = + \sfrac{1}{4} \, , & &\bold{b} = - \sfrac{1}{8} \, , \\
\nonumber
 s_3 &= + \sfrac{1}{4} \, , & &f_3 = - \sfrac{1}{8} \, , & &g_3 = - \sfrac{1}{4} \, , & &\bold{c} = + \sfrac{1}{6} \, , \\
\nonumber
 s_4 &= - \sfrac{1}{8} \, , & &f_4 = + \sfrac{1}{48} \, , & &g_4 = + \sfrac{1}{4} \, , & &\bold{d} = + \sfrac{1}{4} \, , \\
\nonumber
 s_5 &= + \sfrac{1}{24} \, , & &f_5 = 0 \, , & &g_5 = - \sfrac{1}{4} \, , & &\bold{e} = - \sfrac{1}{24} \, , \\
\nonumber
 s_6 &= 0 \, , & &f_6 = - \sfrac{3}{8} \, , & &g_6 = - \sfrac{1}{2} \, , & &\bold{f} = + \sfrac{1}{288} \, , \\
\nonumber
 s_7 &= 0 \, , & &f_7 = - \sfrac{1}{48} \, , & &g_7 = + \sfrac{1}{8} \, , & &\hat{\bold{d}} = + \sfrac{1}{4} \, , \\
\nonumber
 s_8 &= 0 \, , & &f_8 = + \sfrac{1}{8} \, , & &g_8 = 0 \, , & &\hat{\bold{e}} = - \sfrac{1}{4} \, , \\
%
 s_9 &= + \sfrac{1}{4} \, , & &f_9 = - \sfrac{1}{48} \, , & &g_9 =0 \, & &\hat{\bold{f}} = + \sfrac{1}{24} \, , \label{Numerical_Results} \\ 
\nonumber
 s_{10} &= - \sfrac{1}{48} \, , & &f_{10} = + \sfrac{1}{8} \, , & &g_{10} = + \sfrac{1}{8} \, , & &\hat{\bold{g}} = - \sfrac{1}{8} \, , \\
\nonumber
 s_{11} &= + \sfrac{1}{8} \, , & &f_{11} = + \sfrac{1}{288} \, , & &g_{11} = 0 \, , & &\hat{\bold{h}} = + \sfrac{1}{4} \, , \\
\nonumber
 && && &g_{12} = 0 \, , & &\hat{\bold{i}} = - \sfrac{1}{4} \, , \\
\nonumber
 && && &g_{13} = 0 \, , & &\hat{\bold{j}} = - \sfrac{1}{4} \, , \\
\nonumber
 && && &g_{14} = + \sfrac{1}{4} \, , & &\hat{\bold{k}} = + \sfrac{1}{8} \, , \\
\nonumber
 && && &g_{15} = - \sfrac{1}{48} \, , & &\hat{\bold{l}} = - \sfrac{1}{8} \, , \\
 %
%
\nonumber
 && && && &\hat{\bold{n}} = - \sfrac{1}{8} \, , \\  
%
%
\nonumber
 && && && &\hat{\bold{p}} = - \sfrac{1}{16} \, .
\end{align}
%

\subsection{\sl Closure of the Superalgebra} \label{Closure}
We have seen that the higher derivative corrected Euclidean BLG theory is invariant under our supersymmetry ansatz. However, for a truly supersymmetric theory the supersymmetry transformations must close on-shell on to translations and gauge transformations. In this section we show the superalgebra does indeed close for the coefficients listed in \eref{Numerical_Results}. In the absence of cubic fermion terms in $\delta ' X^I_a$ and $\delta ' \tilde{A}_\mu{}^b{}_a$ and quadratic fermions in $\delta ' \psi_a$ we are unable to close on the fermion field. 

We present only our results as the detailed calculations are long. Our methodology in the closure calculations is the same for both the scalar and gauge fields and we detail it here. We first separate out certain terms according to their number of covariant derivatives and then insert the relevant supersymmetry transformations. Next, we use the relation $\{ \Gamma^\mu , \Gamma^I \} = 0$ to group all worldvolume $\Gamma$-matrices together and then expand them out using the Clifford algebra relation. Following this, we perform the $( 1 \leftrightarrow 2 )$ anti-symmetrisation in the supersymmetry parameters making heavy use of \eref{Commutator_Relation}. The transverse $\Gamma$-matrix algebra is performed next and our calculations have again been helped by using the symbolic computer package Cadabra \cite{Peeters:2006kp}\cite{Peeters:2007wn}. Finally, we simplify the remaining expressions wherever possible using the identities in Eqs.\,(\ref{Useful_Id_5}) and (\ref{Useful_Id_4}).
%
\subsubsection{Closure on the Scalar Fields} \label{Scalar_Closure}
The full supersymmetry transformations can be written as $\tilde{\delta} = \delta + \delta'$ where $\delta$ are the lowest order variations and $\delta'$ are the $1/T_{M2}$ corrections. Closure on the scalars then takes the form
\begin{align}
[ \tilde{\delta}_1 , \tilde{\delta}_2 ] X^I_a = [ \delta_1 , \delta_2 ] X^I_a + (\delta_1 \delta_2' + \delta'_1 \delta_2 ) X^I_a - (\delta_2 \delta_1' + \delta'_2 \delta_1 ) X^I_a + [ \delta'_1 , \delta'_2 ] X^I_a \, .
\end{align}
The lowest order commutator, $[ \delta_1 , \delta_2 ] X^I_a$, closes on to translations and gauge transformations \cite{Bagger:2007jr} as we have seen previously. The commutator $[ \delta'_1 , \delta'_2 ] X^I_a$ is $\mathcal{O} (T_{M2}^{-2})$ and can be ignored because we are not considering the $T_{M2}^{-2}$ corrections to the supersymmetry transformations. The remaining mixed terms, $(\delta_1 \delta_2' + \delta'_1 \delta_2 ) X^I_a - (\delta_2 \delta_1' + \delta'_2 \delta_1 ) X^I_a$, are the focus of this section and must be zero for the algebra to close. As closing on the scalar field does not involve use of the equation of motion the mixed terms must be zero either through symmetry arguments or by constraining the coefficients to be zero. Performing the supervariations we find that the resulting terms can be grouped according to the number of covariant derivatives they contain.
To begin, we consider terms which involve three covariant derivatives,
%
\begin{align}
\nonumber
T_{M2} \, (\delta_1 \delta_2' X^I_a + \delta'_1 \delta_2 X^I_a  )_{3DX} &- (1 \leftrightarrow 2) \\
\nonumber
	=& +i ( 6 f_1 - 2 s_1 - 2 s_2 ) ( \bar{\epsilon}_2 \Gamma^{JK} \Gamma^{\mu \nu \lambda} \epsilon_1) D_{\mu} X^{I}_b D_{\nu} X^{J}_c D_{\lambda} X^{K}_d \ d^{bcd}{}_{a} \\
\nonumber	
	&+ i ( 2 f_2 + 2 s_2 - 2 s_3 ) ( \bar{\epsilon}_2 \Gamma^\mu \epsilon_1) D_{\mu} X^{J}_b D_{\nu} X^{I}_c D^{\nu} X^{J}_d \ d^{bcd}{}_{a} \\
	&+ i ( 2 f_3 -  2 s_2 - 2 s_4) ( \bar{\epsilon}_2 \Gamma^{\mu} \epsilon_1) D_{\mu} X^{I}_b D_{\nu} X^{J}_c D^{\nu} X^{J}_d \ d^{bcd}{}_{a} \, .
\end{align}
Closure requires each of these terms is zero. Hence,
\begin{align}
f_1 =& + \sfrac{1}{3} s_1 + \sfrac{1}{3} s_2 \, , \qquad f_2 = - s_2 + s_3 \, , \qquad f_3 = + s_2 + s_4 \, .
\end{align}
%
Next, we consider terms which involve two covariant derivatives,
\begin{align}
\nonumber
T_{M2} \, (\delta_1 \delta_2' X^I_a + \delta'_1 \delta_2 X^I_a  )_{2DX} &- (1 \leftrightarrow 2) \\
\nonumber
=&+ i ( 6 f_4 - s_1 + 2 s_7 ) ( \bar{\epsilon}_2 \Gamma^{JKLM} \Gamma^{\mu \nu} \epsilon_1 ) D_{\mu} X^{J}_b D_{\nu} X^{K}_c X^{ILM}_d \ d^{bcd}{}_{a} \\
\nonumber
&+ i ( 4 f_4 - \sfrac{1}{3} s_2 - 2 s_5 - 2 s_6 ) ( \bar{\epsilon}_2 \Gamma^{JKLM} \Gamma^{\mu \nu} \epsilon_1 ) D_{\mu} X^{I}_b D_\nu X^J_c X^{KLM}_d \ d^{bcd}{}_{a} \\		
\nonumber
&+ i ( 2 f_5 + 2 s_1 - 6 s_5 + 2 s_8 ) ( \bar{\epsilon}_2 \Gamma^{IKLM} \Gamma^{\mu \nu} \epsilon_1 ) D_{\mu} X^{J}_b D_{\nu} X^{K}_c X^{JLM}_d \ d^{bcd}{}_{a} \\		
\nonumber
&+ i ( 2 f_6 + 2 s_1 + 2 s_9 ) ( \bar{\epsilon}_2 \Gamma^{\mu \nu} \epsilon_1 ) D_{\mu} X^{J}_b D_{\nu} X^{K}_c X^{IJK}_d \ d^{bcd}{}_{a} \\
\nonumber
&+ i ( 6 f_7 - s_4 + 2 s_7 ) ( \bar{\epsilon}_2 \Gamma^{JK} \epsilon_1 ) D_{\mu} X^{L}_b D^{\mu} X^{L}_c X^{IJK}_d \ d^{bcd}{}_{a} \\
\nonumber
&+ i ( 2 f_8 - s_3 + 6 s_6 + 2 s_8 ) ( \bar{\epsilon}_2 \Gamma^{KL} \epsilon_1 ) D_{\mu} X^{I}_b D^{\mu} X^{J}_c X^{JKL}_d \ d^{bcd}{}_{a} \\	
&+ i ( - 4 f_8 + 4 s_7 + 2 s_9 ) ( \bar{\epsilon}_2 \Gamma^{KL} \epsilon_1 ) D^\mu X^J_b D_{\mu} X^{L}_c X^{IJK}_d \ d^{bcd}{}_{a} \, .
\end{align}
These two derivatives terms are then zero if
\begin{equation}
f_4 = \sfrac{1}{6} s_1 - \sfrac{1}{3} s_7 \, , \qquad f_4 = \sfrac{1}{12} s_2 + \sfrac{1}{2} s_5 + \sfrac{1}{2} s_6 \, , 
\end{equation}
\begin{equation}
f_5 = - s_1 + 3 s_5 - s_8 \, ,  \qquad f_6 = - s_1 - s_9 \, , \qquad f_7 = \sfrac{1}{6} s_4 - \sfrac{1}{3} s_7 \, ,
\end{equation}
\begin{equation}
f_8 = \sfrac{1}{2} s_3 - 3 s_6 - s_8 \, , \qquad f_8 = s_7 + \sfrac{1}{2} s_9 \, .
\end{equation}
The terms which involve a single covariant derivative are
%
\begin{align}
\nonumber
T_{M2} \, (\delta_1 \delta_2' X^I_a + \delta'_1 \delta_2 X^I_a  )_{1DX} &- (1 \leftrightarrow 2) \\
\nonumber
=&+ i ( 2 f_9 - 2 s_6 - 2 s_{10} ) ( \bar{\epsilon}_2 \Gamma^\mu \epsilon_1 ) D_\mu X^I_b X^{JKL}_c X^{JKL}_d \, d^{bcd}{}_a \\
&+ i ( 2 f_{10} - 2 s_7 - 2 s_8 - 2 s_{11} ) ( \bar{\epsilon}_2 \Gamma^\mu \epsilon_1 ) D_\mu X^J_b X^{IKL}_c X^{JKL}_d \, d^{bcd}{}_a \, .
\end{align}
Closure requires 
\begin{equation}
f_9 = s_6 + s_{10} \, ,  \qquad f_{10} = s_7 + s_8 + s_{11} \, .
\end{equation}
Finally, we consider those terms which contain no covariant derivatives,
%
\begin{align}
T_{M2} \, (\delta_1 \delta_2' X^I_a + \delta'_1 \delta_2 X^I_a  )_{0DX} - (1 \leftrightarrow 2) =&+ i \left( 6 f_{11} - s_{10} - \sfrac{1}{3} s_{11} \right) ( \bar{\epsilon}_2 \Gamma^{JK} \epsilon_1 ) X^{IJK}_b X^{LMN}_c X^{LMN}_d \, d^{bcd}{}_a \, ,
\end{align}
and we require
\begin{equation}
f_{11} = \sfrac{1}{6} s_{10} + \sfrac{1}{18} s_{11} \, .
\end{equation}

It is easily verified that the conditions for closure are satisfied when the $f$ and $s$ coefficients take the values found in \eref{Numerical_Results}.

\subsubsection{Closure on the Gauge Fields} \label{Gauge_Closure}
Closing the algebra on $\tilde{A}_\mu$ gives
 \begin{align}
[ \tilde{\delta}_1 , \tilde{\delta}_2 ] \tilde{A}_\mu{}^b{}_a = [ \delta_1 , \delta_2 ] \tilde{A}_\mu{}^b{}_a + (\delta_1 \delta_2' + \delta'_1 \delta_2 ) \tilde{A}_\mu{}^b{}_a - (\delta_2 \delta_1' + \delta'_2 \delta_1 ) \tilde{A}_\mu{}^b{}_a + [ \delta'_1 , \delta'_2 ] \tilde{A}_\mu{}^b{}_a \, .
\end{align}
As for the scalar field, the terms in $[ \delta'_1 , \delta'_2 ] \tilde{A}_\mu{}^b{}_a$ can be ignored. The lowest order terms may be written as
\begin{align}
[\delta_1,\delta_2] \tilde A_\mu{}^b{}_a =& + 2 i ( \bar{\epsilon}_2 \Gamma^\nu \epsilon_1 ) \varepsilon_{\mu\nu\lambda} \left( \sfrac{1}{2} \varepsilon^{\rho \sigma \lambda} \tilde{F}_{\rho \sigma}{}^b{}_a + E_{A_\lambda{}^a{}_b} \right) - 2i ( \bar{\epsilon}_2 \Gamma_{IJ} \epsilon_1 ) X^I_c D_\mu X^J_d f^{cdb}{}_a \, ,
\end{align}
where $E_{A_\lambda{}^a{}_b}$ is the lowest order gauge field equation of motion. From the presence of covariant derivatives in the higher derivative Lagrangian it follows that the gauge field equation of motion picks up $1/T_{M2}$ corrections. Hence for on-shell closure we require the mixed terms make the following contribution to the higher order equation of motion
\begin{align}
(\delta_1 \delta_2' &+ \delta'_1 \delta_2 ) \tilde{A}_\mu{}^b{}_a - ( 1 \leftrightarrow 2 ) = + \sfrac{2 i}{T_{M2}} ( \bar{\epsilon}_2 \Gamma^\nu \epsilon_1 ) \varepsilon_{\mu\nu\lambda} E'_{A_\lambda{}^a{}_b} \\
%
%
\nonumber
=&+ \sfrac{2 i}{T_{M2}} ( \bar{\epsilon}_2 \Gamma^\nu \epsilon_1 ) \varepsilon_{\mu\nu\lambda} \Big( + 4 \bold{a} \, D^\lambda X^J_e D^\rho X^J_f D_\rho X^I_g X^I_c + 4 \bold{b} D^\lambda X^I_e D^\rho X^J_f D_\rho X^J_g X^I_c \Big) \, d^{efg}{}_d f^{cdb}{}_a \\
\nonumber
&+ \sfrac{2 i}{T_{M2}} ( \bar{\epsilon}_2 \Gamma^\nu \epsilon_1 ) \varepsilon_{\mu\nu\lambda} \Big( + 3 \bold{c} \, \varepsilon^{\rho \sigma \lambda} D_\rho X^J_e D_\sigma X^K_f X^{IJK}_g X^I_c \Big) \, d^{efg}{}_d f^{cdb}{}_a \\
\nonumber
&+ \sfrac{2 i}{T_{M2}} ( \bar{\epsilon}_2 \Gamma^\nu \epsilon_1 ) \varepsilon_{\mu\nu\lambda} \Big( + \left( \sfrac{2}{3} \bold{d} + 2 \bold{e} \right) D^\lambda X^I_e X^{JKL}_f X^{JKL}_g X^I_c \Big) \, d^{efg}{}_d f^{cdb}{}_a \\
&+ \mathcal{O} ( \psi^2 ) \label{Gauge_ReqEOM} \, ,
\end{align}
with all others terms in $(\delta_1 \delta_2' + \delta'_1 \delta_2 ) \tilde{A}_\mu{}^b{}_a - ( 1 \leftrightarrow 2 )$ being zero.
Once again, the closure terms can be neatly split according to their number of covariant derivatives. We first consider terms which involve three covariant derivatives,
%
\begin{align}
\nonumber
T_{M2} \, (\delta_1 \delta_2' \tilde{A}_\mu{}^b{}_a &+ \delta'_1 \delta_2 \tilde{A}_\mu{}^b{}_a )_{3DX} - (1 \leftrightarrow 2) \\
\nonumber
=&+ i ( 2 f_2 - 2 g_5 - 2 g_6 ) \varepsilon_{\mu \nu \lambda} ( \bar{\epsilon}_2 \Gamma^{\nu} \epsilon_1 ) X^I_c D^{\rho} X^{I}_e D^{\lambda} X^{J}_f D_{\rho} X^{J}_g \, d^{efg}{}_{d} f^{cdb}{}_a \\
\nonumber
&+ i ( 2 f_3 - 2 g_1 + 2 g_6 ) \varepsilon_{\mu \nu \lambda} ( \bar{\epsilon}_2 \Gamma^{\nu} \epsilon_1 ) X^I_c D^{\lambda} X^{I}_e D_{\rho} X^{J}_f D^{\rho} X^{J}_g \, d^{efg}{}_{d} f^{cdb}{}_a \\
\nonumber
&+ i ( - 2 g_2 + 2 g_3 - 2 g_6 ) \varepsilon_{\nu \lambda \rho} ( \bar{\epsilon}_2 \Gamma^\nu \epsilon_1 ) X^I_c D^\lambda X^I_e D_\mu X^J_f D^\rho X^J_g \, d^{efg}{}_d f^{cdb}{}_a \\
\nonumber
&+ i ( - 2 g_3 + 2 g_5 - 2 g_8 ) ( \bar{\epsilon}_2 \Gamma^{JK} \epsilon_1 ) D_\mu X^J_e D_\nu X^K_f D^\nu X^I_g X^I_c \, d^{efg}{}_d f^{cdb}{}_a \\			
\nonumber
&+ i ( - 6 f_1 + 2 g_7 + 2 g_8 ) \varepsilon^{\nu \lambda \rho} ( \bar{\epsilon}_2 \Gamma^{IJKL} \Gamma_\rho \epsilon_1 ) X^I_c D_{\mu} X^J_e D_{\nu} X^K_f D_{\lambda} X^L_g \, d^{efg}{}_{d} f^{cdb}{}_a \\	
\nonumber
&+ i ( 2 f_3 - 2 g_1 - 2 g_8 ) ( \bar{\epsilon}_2 \Gamma^{IJ} \epsilon_1 ) X^I_c D_{\mu} X^{J}_e D_{\nu} X^{K}_f D^{\nu} X^{K}_g \, d^{efg}{}_{d} f^{cdb}{}_a \\			
&+ i ( 2 f_2 - 2 g_2 + 2 g_8 ) ( \bar{\epsilon}_2 \Gamma^{IK} \epsilon_1 ) X^I_c D_{\mu} X^{J}_e D_{\nu} X^{J}_f D^{\nu} X^{K}_g \, d^{efg}{}_{d} f^{cdb}{}_a \, .
\end{align}
The first and second terms form part of the higher derivative equation of motion and after comparing with \eref{Gauge_ReqEOM} we find
\begin{equation}
2 f_2 - 2 g_5 - 2 g_6 = 8 \bold{a} \, , \qquad 2 f_3 - 2 g_1 + 2 g_6 = 8 \bold{b} \, .
\end{equation}
The remaining coefficients must be zero for closure of the superalgebra. Hence,
\begin{equation}
f_1 = \sfrac{1}{3} g_7 + \sfrac{1}{3} g_8 \, , \qquad f_2 = g_2 - g_8 \, , \qquad f_3 = g_1 + g_8 \, ,
\end{equation}
\begin{equation}
g_2 - g_3 + g_6 = 0 \, , \qquad  g_3 - g_5 + g_8 =0 \, .
\end{equation}
Next we consider terms which involve two covariant derivatives,
%
%
\begin{align}
\nonumber
T_{M2} \, (\delta_1 \delta_2' \tilde{A}_\mu{}^b{}_a &+ \delta'_1 \delta_2 \tilde{A}_\mu{}^b{}_a )_{2DX} - (1 \leftrightarrow 2) \\
\nonumber
=&+ i ( 4 f_6 + 2 g_8 - 2 g_{11} - 2 g_{14} ) ( \bar{\epsilon}_2 \Gamma^{\nu} \epsilon_1 ) X^I_c D_\mu X^J_e D_\nu X^K_f X^{IJK}_g \, d^{efg}{}_d f^{cdb}{}_a \\	
\nonumber
&+ i ( 2 f_5 + 2 f_6 - g_6 + 2 g_7 + 6 g_9 ) \varepsilon_{\mu \nu \lambda} ( \bar{\epsilon}_2 \Gamma^{KL} \epsilon_1 ) X^I_c D^\nu X^I_e D^\lambda X^J_f X^{JKL}_g \, d^{efg}{}_d f^{cdb}{}_a \\	
\nonumber
&+ i ( 4 f_5 + 4 g_7 - 4 g_{10} + 2 g_{11} ) \varepsilon_{\mu \nu \lambda} ( \bar{\epsilon}_2 \Gamma^{KL} \epsilon_1 ) X^I_c D^\nu X^J_e D^\lambda X^L_f X^{IJK}_g \, d^{efg}{}_d f^{cdb}{}_a \\
\nonumber
&+ i ( 2 f_8 + g_5 - 6 g_9 ) ( \bar{\epsilon}_2 \Gamma_\mu \Gamma^{IJKL} \epsilon_1 ) X^I_c D^\nu X^J_e D_\nu X^M_f X^{KLM}_g \, d^{efg}{}_d f^{cdb}{}_a \\	
\nonumber
&+ i ( 12 f_4 - 2 f_5 + g_3 - g_8 + 6 g_9 ) ( \bar{\epsilon}_2 \Gamma^{\nu} \Gamma^{IJLM} \epsilon_1 ) X^I_c D_\mu X^J_e D_\nu X^K_f X^{KLM}_g \, d^{efg}{}_d f^{cdb}{}_a \\		
\nonumber
&+ i ( - 12 f_4 + 2 f_5 + g_4 + g_8 - 6 g_{12} ) ( \bar{\epsilon}_2 \Gamma^{\nu} \Gamma^{IJKL} \epsilon_1 ) X^I_c D_\nu X^J_e D_\mu X^M_f X^{KLM}_g \, d^{efg}{}_d f^{cdb}{}_a \\	
&+ i ( 12 f_4 - g_8 - 2 g_{10} - 2 g_{13} ) ( \bar{\epsilon}_2 \Gamma^{\nu} \Gamma^{JKLM} \epsilon_1 ) X^I_c D_\mu X^J_e D_\nu X^K_f X^{ILM}_g \, d^{efg}{}_d f^{cdb}{}_a \, .		
\end{align}
The first term contributes to the gauge field equation of motion. After multiplying out the $\varepsilon$-tensors in \eref{Gauge_ReqEOM} we find that closure on-shell requires
\begin{equation}
4 f_6 + 2 g_8 - 2 g_{11} - 2 g_{14} = -12 \bold{c} \, . 
\end{equation}
The remaining terms are zero provided 
\begin{equation}
f_4 = \sfrac{1}{12} g_8 + \sfrac{1}{6} g_{10} + \sfrac{1}{6} g_{13} \, , 
\end{equation}
\begin{equation}
6 f_4 - f_5 = - \sfrac{1}{2} g_3 + \sfrac{1}{2} g_8 - 3 g_9 \, , \qquad 6 f_4 - f_5 = \sfrac{1}{2} g_4 + \sfrac{1}{2} g_8 - 3 g_{12} \, ,
\end{equation}
\begin{equation}
f_5 = - g_7 + g_{10} - \sfrac{1}{2} g_{11} \, ,
\end{equation}
\begin{equation}
f_5 + f_6 = \sfrac{1}{2} g_6 - g_7 - 3 g_9  \, , \qquad f_8 = - \sfrac{1}{2} g_5 + 3 g_9 \, .
\end{equation}
The terms which involve a single covariant derivative are
%
%
\begin{align}
\nonumber
T_{M2} \, (\delta_1 \delta_2' \tilde{A}_\mu{}^b{}_a &+ \delta'_1 \delta_2 \tilde{A}_\mu{}^b{}_a )_{1DX} - (1 \leftrightarrow 2) \\
\nonumber
=&+ i ( 2 f_9 + \sfrac{2}{3} f_{10} + 2 g_9 + \sfrac{2}{3} g_{10} - 2 g_{15} ) ( \bar{\epsilon}_2 \Gamma^{\nu} \epsilon_1 ) \varepsilon_{\mu \nu \lambda} D^\lambda X^I_e X^{JKL}_f X^{JKL}_g X^I_c \, d^{efg}{}_d f^{cdb}{}_a \\	
\nonumber
&+ i ( 2 f_9 - \sfrac{2}{3} g_{13} - 2 g_{15} ) ( \bar{\epsilon}_2 \Gamma^{IJ} \epsilon_1 ) D_\mu X^J_e X^{KLM}_f X^{KLM}_g X^I_c \, d^{efg}{}_d f^{cdb}{}_a \\
\nonumber
&+ i ( 2 f_{10} + 2 g_{13} - g_{14} ) ( \bar{\epsilon}_2 \Gamma^{LM} \epsilon_1 ) D_\mu X^J_e X^{IJK}_f X^{KLM}_g X^I_c \, d^{efg}{}_d f^{cdb}{}_a \, .
\end{align}
The first term forms part of the gauge field equation of motion. Comparing with \eref{Gauge_ReqEOM} we see that
\begin{align}
2 f_9 + \sfrac{2}{3} f_{10} + 2 g_9 + \sfrac{2}{3} g_{10} - 2 g_{15} = + \sfrac{4}{3} \bold{d} + 4 \bold{e} \, .
\end{align}
The remaining coefficients must be zero hence,
\begin{equation}
f_9 = \sfrac{1}{3} g_{13} + g_{15} \, , \qquad f_{10} = - g_{13} + \sfrac{1}{2} g_{14}  \, .
\end{equation}
Next we consider terms which involve no covariant derivatives,
%
\begin{align}
\nonumber
T_{M2} \, (\delta_1 \delta_2' \tilde{A}_\mu{}^b{}_a &+ \delta'_1 \delta_2 \tilde{A}_\mu{}^b{}_a )_{0DX} - (1 \leftrightarrow 2) \\
=&+ i \left( 2 f_{11} - \sfrac{1}{3} g_{15} \right) ( \bar{\epsilon}_2 \Gamma_\mu \Gamma^{IJKL} \epsilon_1 ) X^{JKL}_e X^{MNO}_f X^{MNO}_g X^I_c \, d^{efg}{}_d f^{cdb}{}_a \, .
\end{align}
At first sight we should take the coefficient to be zero however, using the identity \eqref{Useful_Id_4} we can show that the term is zero independently of its coefficient and consequently this part of algebra closes automatically.

Once more it is easy to verify that all the gauge field closure conditions are satisfied by the coefficients listed in \eref{Numerical_Results}.

\subsection{\sl Summary of Results} \label{Summary}
In summary, we have found that the maximally supersymmetric higher derivative corrected Lagrangian of the $\mathcal{A}_4$ BLG theory, to lowest non-trivial order in fermions, is
\begin{align}
\nonumber
\mathcal{L} = \mathcal{L}_{BLG} + \sfrac{1}{T_{M2}} {\rm STr} \Big\{ &+ \sfrac{1}{4} \, D^\mu X^I D_\mu X^J  D^\nu X^J D_\nu X^I - \sfrac{1}{8} \, D^\mu X^I D_\mu X^I D^\nu X^J D_\nu X^J \\
\nonumber
&+ \sfrac{1}{6} \, \varepsilon^{\mu\nu\lambda}\, X^{IJK} D_\mu X^I D_\nu X^J D_\lambda X^K \\
\nonumber
&+ \sfrac{1}{4} \, X^{IJK}X^{IJL} D^\mu X^K D_\mu X^L - \sfrac{1}{24} \, X^{IJK} X^{IJK} D^\mu X^L  D_\mu X^L \\
\nonumber
&+ \sfrac{1}{288} \, X^{IJK}X^{IJK} X^{LMN}X^{LMN} \\
%
%
\nonumber
&+ \sfrac{i}{4} \, \bar{\psi} \Gamma^\mu \Gamma^{IJ} D^\nu\psi {D}_\mu X^{I} {D}_\nu X^J - \sfrac{i}{4} \, \bar{\psi}\Gamma^\mu D^\nu\psi  {D}_\mu X^{I} {D}_\nu X^I \\
\nonumber
&+ \sfrac{i}{24} \, \bar{\psi}\Gamma^{IJKL} D^\nu\psi\; X^{IJK}  {D}_\nu X^L - \sfrac{i}{8} \, \bar{\psi}\Gamma^{IJ}  D^\nu\psi\; X^{IJK}{D}_\nu X^K \\
\nonumber
&+ \sfrac{i}{4} \, \bar{\psi}\Gamma^{IJ}[X^J,X^{K},\psi] {D}^\mu X^{I} {D}_\mu X^K \\
\nonumber
&- \sfrac{i}{4} \, \bar{\psi}\Gamma^{\mu\nu}[X^I,X^{J},\psi] {D}_\mu X^{I}{D}_\nu X^J - \sfrac{i}{4} \, \bar{\psi}\Gamma^{\mu\nu}\Gamma^{IJ}[X^J,X^{K},\psi] {D}_\mu X^{I}{D}_\nu X^K \\
\nonumber
&+ \sfrac{i}{8} \, \bar{\psi}\Gamma^\mu\Gamma^{IJ}[X^K,X^{L},\psi] {D}_\mu X^{I}X^{JKL} - \sfrac{i}{8} \, \bar{\psi}\Gamma^\mu[X^I,X^{J},\psi] {D}_\mu X^{K}X^{IJK} \\
\nonumber
&- \sfrac{i}{8} \, \bar{\psi}\Gamma^\mu\Gamma^{IJ}[X^K,X^{L},\psi] {D}_\mu X^L X^{IJK} \\
&- \sfrac{i}{16} \, \bar{\psi}\Gamma^{IJ} [X^K,X^{L},\psi] X^{IJM}X^{KLM} \Big\} \, .
\end{align}
The preceding Lagrangian is invariant under the following $\mathcal{N}=8$ supersymmetry transformations;
\begin{align}
%
%
\nonumber
\delta X^I_a = i ( \bar{\epsilon} \Gamma^{I} \psi_a ) + \sfrac{1}{T_{M2}} \big\{ &+\sfrac{i}{8} ( \bar{\epsilon} \Gamma^{IJK} \Gamma^{\mu \nu} \psi_b ) D_{\mu} X^{J}_c D_{\nu} X^{K}_d \ d^{bcd}{}_{a} \\
\nonumber
&+ \sfrac{i}{4} ( \bar{\epsilon} \Gamma^{J} \psi_b ) D_{\mu} X^{I}_c D^{\mu} X^{J}_d \ d^{bcd}{}_{a} \\
\nonumber
&- \sfrac{i}{8} ( \bar{\epsilon} \Gamma^{I} \psi_b ) D_{\mu} X^{J}_c D^{\mu} X^{J}_d \ d^{bcd}{}_{a} \\ 
\nonumber
&+ \sfrac{i}{24} ( \bar{\epsilon} \Gamma^{IJKLM} \Gamma^{\mu} \psi_b ) D_{\mu} X^{J}_c X^{KLM}_d \ d^{bcd}{}_{a} \\
\nonumber
&+ \sfrac{i}{4} ( \bar{\epsilon} \Gamma^{M} \Gamma^{\mu} \psi_b ) D_{\mu} X^{J}_c X^{IJM}_d \ d^{bcd}{}_{a} \\
\nonumber
&- \sfrac{i}{48} ( \bar{\epsilon} \Gamma^{I} \psi_b ) X^{JKL}_c X^{JKL}_d \ d^{bcd}{}_{a} \\
&+ \sfrac{i}{8} ( \bar{\epsilon} \Gamma^{L} \psi_b ) X^{JKL}_c X^{JKI}_d \ d^{bcd}{}_{a} \big\} \, , \\
\nonumber \\
%
%
%
%
\nonumber
\delta \psi_a =\Gamma^\mu \Gamma^I \epsilon \, D_\mu X^I_a - \sfrac{1}{6} \Gamma^{IJK} \epsilon \, X^{IJK}_a + \sfrac{1}{T_{M2}} \big\{ &+ \sfrac{1}{24} \Gamma^{JKL} \Gamma^{\mu \nu \lambda} \epsilon \, D_{\mu} X^{J}_b D_{\nu} X^{K}_c D_{\lambda} X^{L}_d \ d^{bcd}{}_{a} \\
\nonumber
&+ \sfrac{1}{4} \Gamma^{K} \Gamma^{\mu} \epsilon \, D_{\mu} X^{J}_b D_{\nu} X^{J}_c D^{\nu} X^{K}_d \ d^{bcd}{}_{a} \\
\nonumber
&- \sfrac{1}{8} \Gamma^{K} \Gamma^{\mu} \epsilon \, D_{\mu} X^{K}_b D_{\nu} X^{J}_c D^{\nu} X^{J}_d \ d^{bcd}{}_{a} \\
\nonumber
&+ \sfrac{1}{48} \Gamma^{JKLMN} \Gamma^{\mu \nu} \epsilon \, D_\mu X^J_b D_\nu X^K_c X^{LMN}_d \, d^{bcd}{}_a \\
\nonumber
&- \sfrac{3}{8} \Gamma^M \Gamma^{\mu \nu} \epsilon \, D_\mu X^J_b D_\nu X^K_c X^{JKM}_d \, d^{bcd}{}_a \\
\nonumber
&- \sfrac{1}{48} \Gamma^{KLM} \epsilon \, D_\mu X^J_b D^\mu X^J_c X^{KLM}_d \, d^{bcd}{}_a \\
\nonumber
&+ \sfrac{1}{8} \Gamma^{KLM} \epsilon \, D_\mu X^J_b D^\mu X^K_c X^{JLM}_d \, d^{bcd}{}_a \\
\nonumber
&- \sfrac{1}{48} \Gamma^J \Gamma^\mu \epsilon \, D_\mu X^J_b X^{KLM}_c X^{KLM}_d \, d^{bcd}{}_a \\
\nonumber
&+ \sfrac{1}{8} \Gamma^M \Gamma^\mu \epsilon \, D_\mu X^J_b X^{JKL}_c X^{KLM}_d \, d^{bcd}{}_a \\
&+ \sfrac{1}{288} \Gamma^{NOP} \epsilon \, X^{JKL}_b X^{JKL}_c X^{NOP}_d \, d^{bcd}{}_a \big\} \, , \\
\nonumber \\
%
%
%
%
\nonumber
\delta \tilde{A}_{\mu}{}^b{}_a =i \bar{\epsilon} \Gamma_\mu \Gamma_I X^I_c \psi_d f^{cdb}{}_a + \sfrac{1}{T_{M2}} \big\{ &- \sfrac{i}{8} ( \bar{\epsilon} \Gamma_\mu \Gamma^I \psi_e ) D_\nu X^J_f D^\nu X^J_g X^I_c \, d^{efg}{}_d f^{cdb}{}_a \\
\nonumber
&+ \sfrac{i}{4} ( \bar{\epsilon} \Gamma^\nu \Gamma^I \psi_e ) D_\mu X^J_f D_\nu X^J_g X^I_c \, d^{efg}{}_d f^{cdb}{}_a \\
\nonumber
&- \sfrac{i}{4} ( \bar{\epsilon} \Gamma^\nu \Gamma^J \psi_e ) D_\mu X^J_f D_\nu X^I_g X^I_c \, d^{efg}{}_d f^{cdb}{}_a \\
\nonumber
&+ \sfrac{i}{4} ( \bar{\epsilon} \Gamma^\nu \Gamma^J \psi_e ) D_\mu X^I_f D^\nu X^J_g X^I_c \, d^{efg}{}_d f^{cdb}{}_a \\
\nonumber
&- \sfrac{i}{4} ( \bar{\epsilon} \Gamma_\mu \Gamma^J \psi_e ) D_\nu X^J_f D^\nu X^I_g X^I_c \, d^{efg}{}_d f^{cdb}{}_a \\
\nonumber
&- \sfrac{i}{2} ( \bar{\epsilon} \Gamma_{\mu \nu \lambda} \Gamma^J \psi_e ) D^\nu X^J_f D^\lambda X^I_g X^I_c \, d^{efg}{}_d f^{cdb}{}_a \\
\nonumber
&+ \sfrac{i}{8} ( \bar{\epsilon} \Gamma_{\mu \nu \lambda} \Gamma^{IJK} \psi_e ) D^\nu X^J_f D^\lambda X^K_g X^I_c \, d^{efg}{}_d f^{cdb}{}_a \\
\nonumber
&+ \sfrac{i}{8} ( \bar{\epsilon} \Gamma_{\mu \nu} \Gamma^{JLM} \psi_e ) D^\nu X^J_f X^{ILM}_g X^I_c \, d^{efg}{}_d f^{cdb}{}_a \\
\nonumber
&+ \sfrac{i}{4} ( \bar{\epsilon} \Gamma^{M} \psi_e ) D_\mu X^J_f X^{IJM}_g X^I_c \, d^{efg}{}_d f^{cdb}{}_a \\
&- \sfrac{i}{48} ( \bar{\epsilon} \Gamma_{\mu} \Gamma^{I} \psi_e ) X^{JKL}_f X^{JKL}_g X^I_c \, d^{efg}{}_d f^{cdb}{}_a \big\} \, .
\end{align}

\section{\sl Conclusions and Outlook} \label{conclusions}

In this paper we have determined the four-derivative order corrections to both the supersymmetry transformations and Lagrangian of the $\mathcal{A}_4$ Bagger-Lambert-Gustavsson theory. Supersymmetric invariance of the Lagrangian requires that the arbitrary coefficients in the system are fixed up to an overall scale parameter and by reference to the abelian DBI action for a single M2-brane, the scale parameter is itself fixed leading to definite numerical values for all the coefficients. We have also shown that the supersymmetry algebra closes on-shell on to the scalar and gauge fields at linear order in the fermions.\footnote{With the coefficients we have determined, it can also be demonstrated that the presence of higher derivative corrections in the fermion supersymmetry does not modify the BPS equation \cite{Bagger:2006sk}: $\sfrac{dX^a}{d(x^2)} = - \sfrac{i}{6} \varepsilon^{abcd} X^{bcd}$.}

In establishing these results we have made use of the identity $f^{[abcd} f^{e] fgh} =0$, which is trivially satisfied by the structure constants of the $\mathcal{A}_4$ 3-algebra as $f^{abcd} \propto \epsilon^{abcd}$ and $a,b,c,d \in \{1,2,3,4\}$. However, the Lorentzian and other non-Euclidean 3-algebras of \cite{Gomis:2008uv}-\cite{deMedeiros:2009hf} do not necessarily satisfy this identity and it is clear that our results do not hold for these wider classes of theories. Therefore, to extend our results to the non-Euclidean BLG theories we must abandon use of the identities which follow from $f^{[abcd} f^{e] fgh} =0$ i.e.\ Eqs.\,(\ref{Useful_Id}), \eqref{Useful_Id2}, \eqref{Useful_Id_5} and \eqref{Useful_Id_4}. Consequently, we should reinstate the $\bold{g}$, $\hat{\bold{m}}$ and $\hat{\bold{o}}$ terms in $\mathcal{L}_{1/T_{M2}}$ \eqref{L_Higher_Starting_Point} as well as adding terms to the order $1/T_{M2}$ supersymmetry transformations. The coefficients of the new terms would then be determined by repeating the analysis in this paper. We hope to report on the extension of this work to real, non-Euclidean 3-algebras in the future.

This work is incomplete in the sense that we have only worked to lowest order in fermions.  The quartic fermion terms in the action, which coincide for the Lorentzian \cite{Alishahiha:2008rs} and Euclidean \cite{Ezhuthachan:2009sr} BLG theories, are known. Incorporating higher fermions in the supersymmetry transformations would, in principle, allow us to verify that the entire theory at $\mathcal{O} ( 1/T_{M2} )$ is maximally supersymmetric and additionally, to close the superalgebra on all the fields. To proceed at this level would require the addition of supersymmetry transformations of the form
\begin{align}
T_{M2} \, \delta' X =&+ ( \bar{\epsilon} \Gamma \psi ) ( \bar{\psi} \Gamma D \psi ) + ( \bar{\epsilon} \Gamma \psi ) \, \bar{\psi} \Gamma [ \psi , X , X ] \, , \label{X_susy_f} \\[6pt]
T_{M2} \, \delta' \tilde{A} =& + ( \bar{\epsilon} \Gamma \psi ) ( \bar{\psi} \Gamma D \psi ) X + ( \bar{\epsilon} \Gamma \psi ) \, \bar{\psi} \Gamma [ \psi , X , X ] X \, , \label{A_susy_f} \\[6pt]
\nonumber
T_{M2} \, \delta' \psi =&+ \Gamma \epsilon ( \bar{\psi} \Gamma D \psi ) DX + \Gamma \epsilon ( \bar{\psi} \Gamma D \psi ) [X,X,X] \\
&+ \Gamma \epsilon \, \bar{\psi} \Gamma [ \psi ,X,X] DX + \Gamma \epsilon \, \bar{\psi} \Gamma [ \psi ,X,X] [X,X,X] \, . \label{psi_susy_f}
\end{align}
The most general starting point would then involve taking all independent Lorentz invariant combinations. However, the presence of two sets of $\Gamma$-matrices in the supersymmetries allows for many ways of contracting Lorentz indices and also brings into play the transverse duality relation $\star \Gamma^{(n)} \propto \varepsilon_{(8)} \Gamma^{(8-n)}$. In addition, the cubic fermions in Eqs.\,(\ref{X_susy_f}) and (\ref{A_susy_f}) can be rearranged using the Fierz relation. The impact of these features is to obscure which terms are independent so that even the starting point is difficult to determine. Moreover, the subsequent invariance and closure calculations would involve heavy use of the Fierz rearrangement and consequently represent a formidable computational challenge which we leave for the time being.

In certain circumstances, the Euclidean BLG theory has a spacetime interpretation of describing two M2-branes. The theory which describes $N$ M2-branes is the ABJM theory \cite{Aharony:2008ug} with gauge group $U(N) \times U(N)$. This theory has manifest $\mathcal{N}=6$ supersymmetry\footnote{This is enhanced to $\mathcal{N}=8$ when the Chern-Simons level takes the values $k=1,2$ \cite{Aharony:2008ug}\cite{Gustavsson:2009pm}.} together with an $SU(4)$ R-symmetry and can be formulated using complex 3-algebras \cite{Bagger:2008se}. With the exception of the abelian $U(1) \times U(1)$ theory \cite{Sasaki:2009ij}, the order $1/T_{M2}$ higher derivative extension of the ABJM model has not been examined. 

Possible methods of approaching the ABJM higher derivative extension have been discussed in \cite{Ezhuthachan:2009sr} and \cite{Low:2010ie}. A separate brute force approach is simply to consider the most general action and supervariations which are consistent with all symmetries of the system and try to demonstrate invariance and closure as we have done here for the $\mathcal{A}_4$ BLG theory. It is conceivable that the arbitrary coefficients in the $1/T_{M2}$ extension of ABJM can likewise be determined up to an overall scaling parameter. It would then remain to fix this scale parameter and there are at least two possible ways of doing this. First, we could directly compare against multiple D2-branes written in a suitable complex format by using the `novel Higgs mechanism' for ABJM \cite{Li:2008ya}\cite{Pang:2008hw} or perhaps by taking an infinite periodic array of M2-branes \cite{Jeon:2012fn}. Secondly we could re-write the results of this paper in complex $SU(2) \times SU(2)$ form \cite{VanRaamsdonk:2008ft} and exploit the equivalence, at levels $k=1$ and 2, of the $U(2) \times U(2)$ ABJM and $\mathcal{A}_4$ BLG theories. 

\section*{\sl Acknowledgements}

We thank Neil Lambert and Costis Papageorgakis for insightful discussions. We are also grateful to the Isaac Newton Institute for Mathematical Sciences, Cambridge and the organisers of the ``Mathematics and Applications of Branes in String and M-theory" programme for providing a stimulating environment in which to work. PR is supported by the STFC studentship grant ST/F007698/1.

\begin{appendices}

\section{Conventions and Useful Identities} \label{Conventions}

All spinorial quantities are those of the eleven-dimensional Clifford algebra with mostly plus metric, and are taken to be real. We denote the M2-brane worldvolume indices by $\mu, \nu , \ldots = 0, 1, 2$ and transverse indices by $I,J, \ldots = 3, 4, \ldots, 10$. The unbroken supersymmetry parameters, which are 16 component Majorana-Weyl spinors, satisfy the following chirality conditions
\begin{align}
\Gamma_{012} \epsilon =& + \epsilon \, , \\
\Gamma_{012} \psi =& - \psi \, .
\end{align}
From the chirality conditions and the choice $\varepsilon_{012} = - 1$, we deduce the following M2-brane worldvolume $\Gamma$-matrix duality relations
\begin{equation}
\Gamma_{\mu \nu \lambda} \epsilon = - \varepsilon_{\mu \nu \lambda} \epsilon \, , \qquad \Gamma_{\nu \lambda} \epsilon = - \varepsilon_{\mu \nu \lambda} \Gamma^\mu \epsilon \, , \qquad \Gamma_{\lambda} \epsilon = + \sfrac{1}{2} \varepsilon_{\mu \nu \lambda} \Gamma^{\mu \nu} \epsilon \, , \label{wvol_duality_e}
\end{equation}
\begin{equation}
\Gamma_{\mu \nu \lambda} \psi = + \varepsilon_{\mu \nu \lambda} \psi \, , \qquad \Gamma_{\nu \lambda} \psi = + \varepsilon_{\mu \nu \lambda} \Gamma^\mu \psi \, , \qquad \Gamma_{\lambda} \psi = - \sfrac{1}{2} \varepsilon_{\mu \nu \lambda} \Gamma^{\mu \nu} \psi \, . \label{wvol_duality_f}
\end{equation}
%
%
The $\Gamma$-matrices have transposes given by;
\begin{align}
\left( \Gamma^{(\mu_n)} \right)^{\rm T} &= (-1)^{\frac{1}{2} n ( n + 1) } C \Gamma^{(\mu_n)} C^{-1} =
\begin{cases} 
- C \Gamma^{(\mu_n)} C^{-1} \ \text{ if $n=1$ or 2,} \\
+ C \Gamma^{(\mu_n)} C^{-1} \ \text{ if $n=3$,} \label{Transpose_WVol}
\end{cases} \\
\nonumber \\
\left( \Gamma^{(I_m)} \right)^{\rm T} &= (-1)^{\frac{1}{2} m ( m + 1) } C \Gamma^{(I_m)} C^{-1} = 
\begin{cases} 
- C \Gamma^{(I_m)} C^{-1} \ \text{ if $m=1,2,5$ or 6,} \\
+ C \Gamma^{(I_m)} C^{-1} \ \text{ if $m=3, 4, 7$ or 8,} \label{Transpose_Trans}
\end{cases}
\end{align}
where $C=\Gamma_0$ is the anti-symmetric charge conjugation matrix and we denote by $\Gamma^{(\mu_n)}$ and $\Gamma^{(I_m)}$ the totally anti-symmetric product of $n$ worldvolume and $m$ transverse $\Gamma$-matrices respectively. Using the transpose properties \eqref{Transpose_WVol} and \eqref{Transpose_Trans} together with $\{ \Gamma^I , \Gamma^\mu \} =0$ we find for any two spinors $\chi$ and $\lambda$
\begin{align}
\bar{\chi} \Gamma^{(I_m)} \Gamma^{(J_n)} \Gamma^{(\mu_p)} \lambda =& (-1)^{\theta ( m,n,p )} \bar{\lambda} \Gamma^{(J_n)} \Gamma^{(I_m)} \Gamma^{(\mu_p)} \chi \, , \label{Spinor_Flip}
\end{align}
where $\theta ( m,n,p ) = p(m+n) + \frac{1}{2} m ( m + 1 ) + \frac{1}{2} n ( n + 1 ) + \frac{1}{2} p ( p + 1 )$. Hence for $\chi=\epsilon_2$ and $\lambda=\epsilon_1$
\begin{align}
\bar{\epsilon}_2 \Gamma^{(I_m)} \Gamma^{(J_n)} \Gamma^{(\mu_p)} \epsilon_1 - ( 1 \leftrightarrow 2 ) =
\begin{cases}
\bar{\epsilon}_2 \left[ \Gamma^{(I_m)} , \Gamma^{(J_n)} \right] \Gamma^{(\mu_p)} \epsilon_1 \ \text{ if $(-1)^{\theta ( m,n,p )}=+1$,} \\[8pt]
\bar{\epsilon}_2 \left\{ \Gamma^{(I_m)} , \Gamma^{(J_n)} \right\} \Gamma^{(\mu_p)} \epsilon_1 \ \text{ if $(-1)^{\theta ( m,n,p )}=-1$.} \label{Commutator_Relation}
\end{cases}
\end{align}

\end{appendices}


\end{document}